\documentclass[12pt]{iopart}

\usepackage{graphicx}
\usepackage{braket}
\usepackage[caption=false]{subfig}
\usepackage[]{color}

\newcommand{\non}{\nonumber}
\newcommand{\bea}{\begin{eqnarray}}
\newcommand{\eea}{\end{eqnarray}}
\newcommand{\beq}{\begin{equation}}
\newcommand{\eeq}{\end{equation}}

\begin{document}

\title{Quench dynamics and parity blocking in Majorana wires}

\author{Suraj Hegde$^1$, Vasudha Shivamoggi$^2$, Smitha Vishveshwara$^1$,
and Diptiman Sen$^3$}

\address{$^1$Department of Physics, University of Illinois at
Urbana-Champaign, Urbana, Illinois 61801-3080, USA \\
$^2$Northrop Grumman Electronic Systems, Linthicum Heights, Maryland 21090,
USA \\
$^3$Centre for High Energy Physics, Indian Institute of Science, Bengaluru
560 012, India}
\vspace{10pt}


\begin{abstract}
We theoretically explore quench dynamics in a finite-sized topological fermionic $p$-wave superconducting wire with the goal of demonstrating that topological order can have marked effects on such non-equilibrium dynamics. In the case studied here, topological order is reflected in the presence of two (nearly) isolated Majorana fermionic end bound modes together forming an electronic state that can be occupied or not, leading to two (nearly) degenerate ground states characterized by fermion parity. Our study begins with a characterization of the static properties of the finite-sized wire, including the behavior of the Majorana end modes and the form of the tunnel coupling between them; a transfer matrix approach to analytically determine the locations of the zero energy contours where this coupling vanishes; and a Pfaffian approach to map the ground state parity in the associated phase diagram. We next study the quench dynamics resulting from initializing the system in a topological ground state and then dynamically tuning one of the parameters of the Hamiltonian. For this, we develop a dynamic quantum many-body
technique that invokes a Wick's theorem for Majorana fermions, vastly reducing the numerical effort given the exponentially large Hilbert space. We investigate the salient and detailed features of two dynamic quantities - the overlap between the time-evolved state and the instantaneous ground state (adiabatic fidelity) and the residual energy. When the parity of the instantaneous ground state flips successively with time, we find that the time-evolved state can dramatically switch back and forth between this state and an excited state even when the quenching is very slow, a phenomenon that we term ``parity blocking". This parity blocking becomes prominently manifest as non-analytic jumps as a function of time in both dynamic quantities.
\end{abstract}


\maketitle

\section{Introduction}
\label{sec:Introduction}

Of late, two different concepts in quantum many-body theory have elicited a surge of active research, partly stemming
from experimental advances in condensed matter and cold atomic systems – the concepts of quench dynamics ~\cite{Dziarmaga10,Polkovnikov11,Dutta10a,Kibble76,Kibble80,Zurek85,Zurek96,Dziarmaga05,Damski05,Damski06,Polkovnikov05,Polkovnikov08,
Calabrese05,Calabrese06,Cherng06,Mukherjee07,Divakaran08,Deng08,Divakaran09,Mukherjee10,Sengupta08,Mondal08,Sen08,Mondal09,DeGrandi08,Barankov08,
DeGrandi10,Patane08,Patane09,Bermudez09,Bermudez10,Perk09,Sen10,DeGottardi11,Pollmann10,Dutta10b,Hikichi10,Chandran12,Chandran13,Patel13,Mostame14,
Foster14,Dorner05,Rigol14,Canovi14,Mitra12,Iucci09,Flesch08,Khatami12,Gramsch12,He13}
 and topological order ~\cite{Prange87,Hasan10,Qi11,Bernevig2013,Alicea12,leijnse12}.
 Quenching, or ramping, concerns initializing a system in its equilibrium configuration at some point in parameter space followed by inducing non-equilibrium behavior via dynamic tuning of one of the parameters. When
the tuning occurs through a critical point separating two phases of matter, no matter how slow the tuning rate 1/$\tau$, the diverging time scale associated with the critical point and critical exponent $z$ always results in out-of-equilibrium dynamics in its vicinity. The quantum version of such Kibble-Zurek physics, initially studied as a thermal quench during the formation of the early universe ~\cite{Kibble76,Kibble80,Zurek85,Zurek96}, offers a mine of valuable information about the critical point in question.

 In the realm of topological systems, while quantum Hall systems have been hailed for their topological properties for over three decades ~\cite{Prange87}, the recent attention on other systems has also been spectacular ~\cite{Hasan10,Qi11,Bernevig2013,Alicea12,leijnse12}. On the theoretical front, among others, two paradigm low-dimensional models have been avidly studied for their topological properties – the Majorana wire proposed by Kitaev ~\cite{Kitaev01,Franz13}, which is effectively a lattice version of a spinless $p$-wave superconducting wire, and the two-dimensional Kitaev honeycomb model ~\cite{Kitaev06}.

Here we explore the synergy of these two concepts, namely quench dynamics ~\footnote{While the terms quench and ramp
are sometimes used to distinguish between instantaneous change of parameter versus a time-dependent change at some
given rate, respectively, here we use quench in a more general sense to encompass all such dynamic tuning. Our actual studies
 are restricted to the ramp case.} and topological order, a rich study that is still in its infancy. Such a marriage is exciting from at least two perspectives – can quench dynamics act as a probe for topological order? Can the existence of topological order lead to a different realm in non-equilibrium dynamics? Studies of such synergy ~\cite{Bermudez09,Rajak14,Bermudez10,Chandran12,Chandran13} have just begun to explore diverse and exciting phenomena with regards to dynamic evolution of topological features. In previous work by two of the authors of
this article and co-workers, Ref.~\cite{Kells14a}, the term `topological blocking' was coined with regards to the role played by a highlighting feature
of topological order - ground state degeneracies - in quench dynamics. It is known that a system can have several topological sectors which are associated with these degeneracies and are distinguished from each other by an invariant based on a discrete global symmetry~\cite{Kells08,Kells09}. In quenching between a topological and non-topological phase, if the ground states in the two phases belong to different topological sectors and the Hamiltonian commutes with the global symmetry at all times, the system
 never reaches the ground state of the final phase. As a result, the usual expectation that if the quench is sufficiently slow
 (i.e., almost adiabatic), the system always remains in the instantaneous ground state is violated. This topological blocking effect was demonstrated in Ref.~\cite{Kells14a} for the Majorana wire and the Kitaev honeycomb model constrained to
periodic boundary conditions.

The goal here is to explore quench dynamics in a finite-sized Majorana wire having open boundary conditions, a system that has come into the limelight for topological features that we expect to affect dynamics in a profound way. These features concern the presence of isolated zero energy Majorana fermion bound states at the wire ends within the topological phase; their possible experimental detection in the context of spin-orbit coupled wires ~\cite{Oreg10,Lutchyn10,Mourik12,Deng12,Das12,Finck13} has garnered much attention in terms of fundamental physics as well as
 implications for topological quantum computation~\cite{Nayak08}. In the thermodynamic limit, these Majorana end modes together form a Dirac fermion state that can either be occupied or empty. Thus, degenerate topological sectors are identified by fermion parity. In the finite-sized system, tunnel coupling between these end modes splits the degeneracy in a manner that can be tuned by changing
the parameters of the system. Here we explore the quench dynamics of tuning through a succession of parity flips of the ground state. The investigation involving open boundaries requires the formulation of new dynamic quantum many-body techniques, which we develop here. By investigating measures commonly studied in quench dynamics, we demonstrate that topological order drastically affects non-equilibrium behavior, the most dramatic signature stemming from quench-dependent switching of topological sectors.

Our presentation is as follows. In Sec.~\ref{sec:blocking}, we present a brief description of the salient features of topological blocking and the highlights of this work, including our results regarding parity switching and blocking in
a finite-sized Majorana wire. In Sec.~\ref{sec:pwave}, we begin our detailed exposition by reviewing the $p$-wave superconducting wire given by Kitaev's lattice Hamiltonian. We outline a derivation of its bulk spectrum and description of its topological phase diagram based on the presence or absence of Majorana end modes.
 In Sec.~\ref{sec:finitesize}, based on a transfer matrix formalism, we analyze the fate of the Majorana end modes for a finite-sized lattice. We obtain an approximate form for the degeneracy splitting and exact solutions for contours in the phase diagram where the splitting vanishes. We then invoke Kitaev's argument based on Pfaffian methods to determine the ground state parity of the system and confirm that parity switches occur at these degeneracy points. In Sec.~\ref{sec:quenching}, we begin our discussion of the quench dynamics associated with varying a parameter of the underlying Hamiltonian linearly in time.
We summarize the known results in the case of periodic boundary conditions, in particular, the Kibble-Zurek scaling of the post-quench excitations and the topological blocking phenomenon. In Sec.~\ref{sec:realspace}, we describe the real space time-dependent formalism that we use to study this problem numerically. We focus on two measures, namely the overlap between the time evolved state and the instantaneous ground state, which we refer to as adiabatic fidelity as in previous work~\cite{Polkovnikov11}, and the residual energy, which is the difference between the expectation value of the Hamiltonian in the time evolved state and the instantaneous ground state energy. In Sec.~\ref{sec:results}, we extensively discuss our numerical results in detail, pinpointing the effect of topological blocking associated with multiple parity switches. In Sec.~\ref{sec:discussion}, we present an overview of our study and connect it with related phenomena, such as the fractional Josephson effect in junctions of Majorana wires, as well as to experiments.

\section{Topological/Parity blocking in quenching dynamics}
\label{sec:blocking}

\begin{figure}[]
\centering
\includegraphics[width=0.5\textwidth]{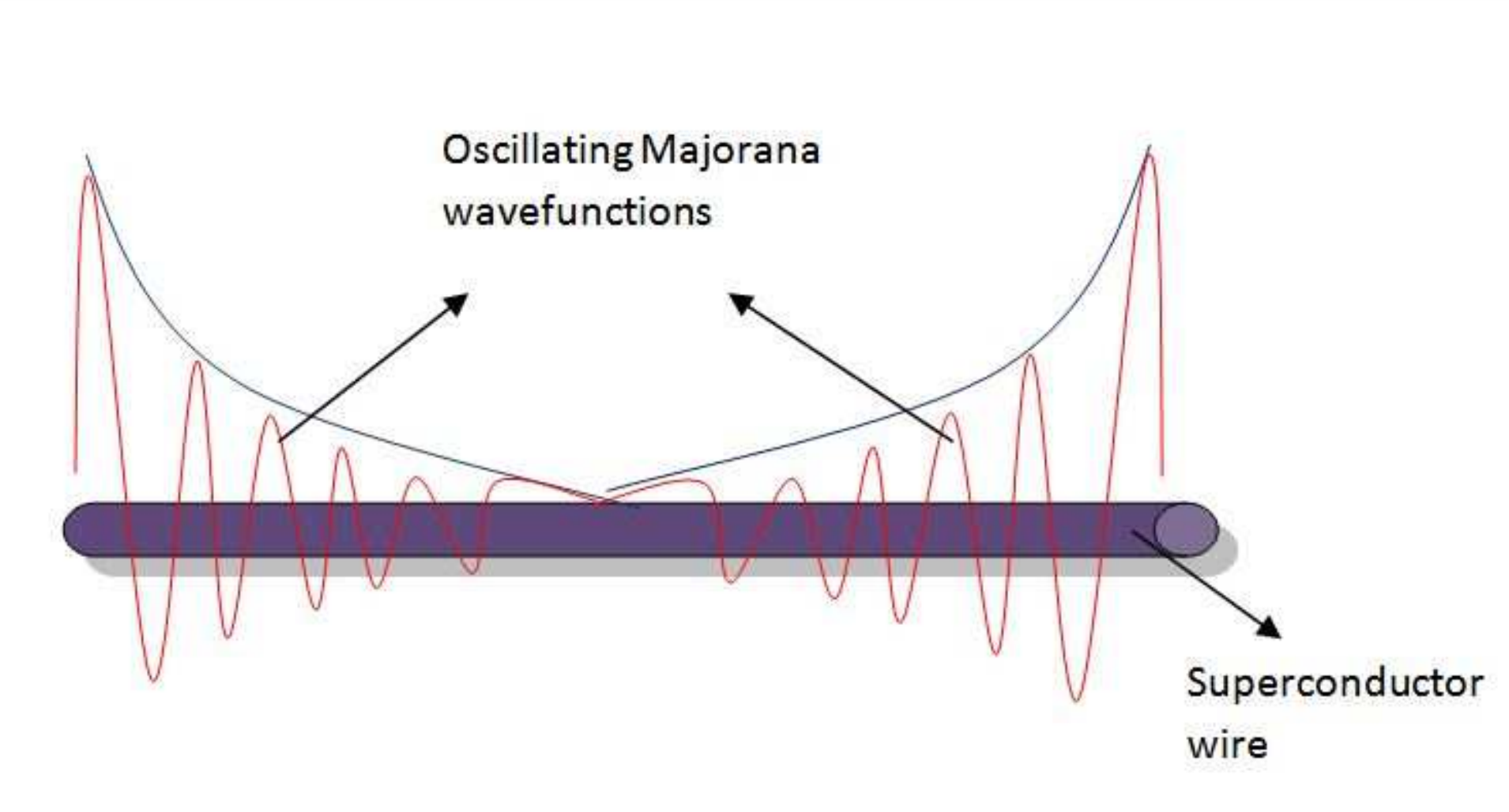}
\caption{(Color online) A finite size superconducting wire in the topological phase characterized
by Majorana fermionic end modes. While their wave functions may or may not oscillate
(red curves), they all decay into the bulk (envelope) over a characteristic length scale
that depends on system parameters. The overlap between the decaying oscillatory
wave functions of these two end modes gives rise to a tunnel splitting of
the otherwise doubly degenerate zero energy states.}\label{fig:cartoon}
\end{figure}

Here, we describe the essence of the topological features that we target with
regards to quench dynamics and present the highlights of this work before
embarking on a detailed exposition.

The stage is set by the concept of topological blocking, which, as mentioned in
the previous section, was studied in Ref.~\cite{Kells14a}. The study
involved quench dynamics in topological systems elicited by changing a parameter
of the Hamiltonian to tune from one quantum phase to another. In going
between a topological phase and a trivial phase, as in the Majorana
wire, or between two topological phases, as in the Kitaev honeycomb model, the focus was
on mismatch of degeneracies. It was shown that if the system was initialized in the ground state
in a phase with higher degeneracy and tuned to one with a lower degeneracy,
two to one in the former
case, and four to three in the latter, the phenomenon of topological blocking would occur.
Certain topological sectors characterized by topological quantum numbers, for instance,
fermion parity, would inhabit the ground state in the initial phase but would have no partner
in the ground state of the final phase. As a result, in tuning through a
quantum critical point separating the two phases, these sectors would evolve so as to have null overlap
with the final ground state no matter how slow the tuning rate, in stark contrast with Kibble-Zurek physics, where only a rate-dependent fraction of the time-evolved state overlaps with the excitation spectrum above the (gapped) final ground state. Moreover, it was shown that even if one took overlap
with the instantaneous sectoral ground state, different topological sectors would show quantitatively
different dynamic behavior, particularly in wave function overlap.

Here, we explore this notion of topological blocking with regards to a different but related aspect - the switching of topological sectors due to quench dynamics within a topological phase. In a Majorana wire with open boundaries, the topological degeneracy is associated with the presence of Majorana zero modes at the edges. The degeneracy is split due finite-size coupling of these edge modes, which induces fermionic parity sectors within the topological region of the phase diagram. Here we build on the notion of topological/parity blocking arising from tuning through these parity sectors.

\begin{figure*}[htp]
\centering
\subfloat[][Adiabatic fidelity for $N=34$.]{\includegraphics[width=0.5\textwidth]{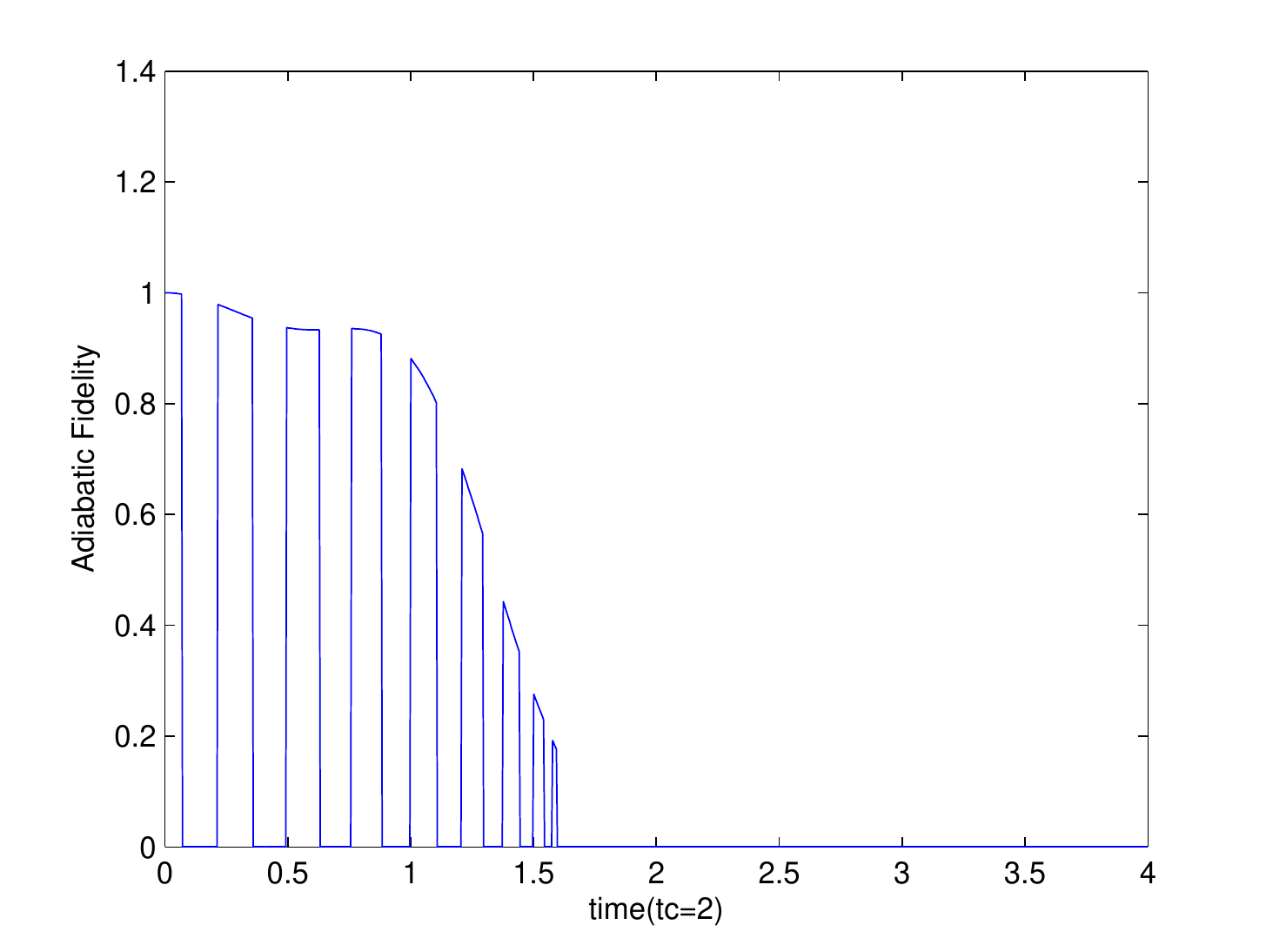}}
\subfloat[][Parity for $N=34$.]{\includegraphics[width=0.5\textwidth]{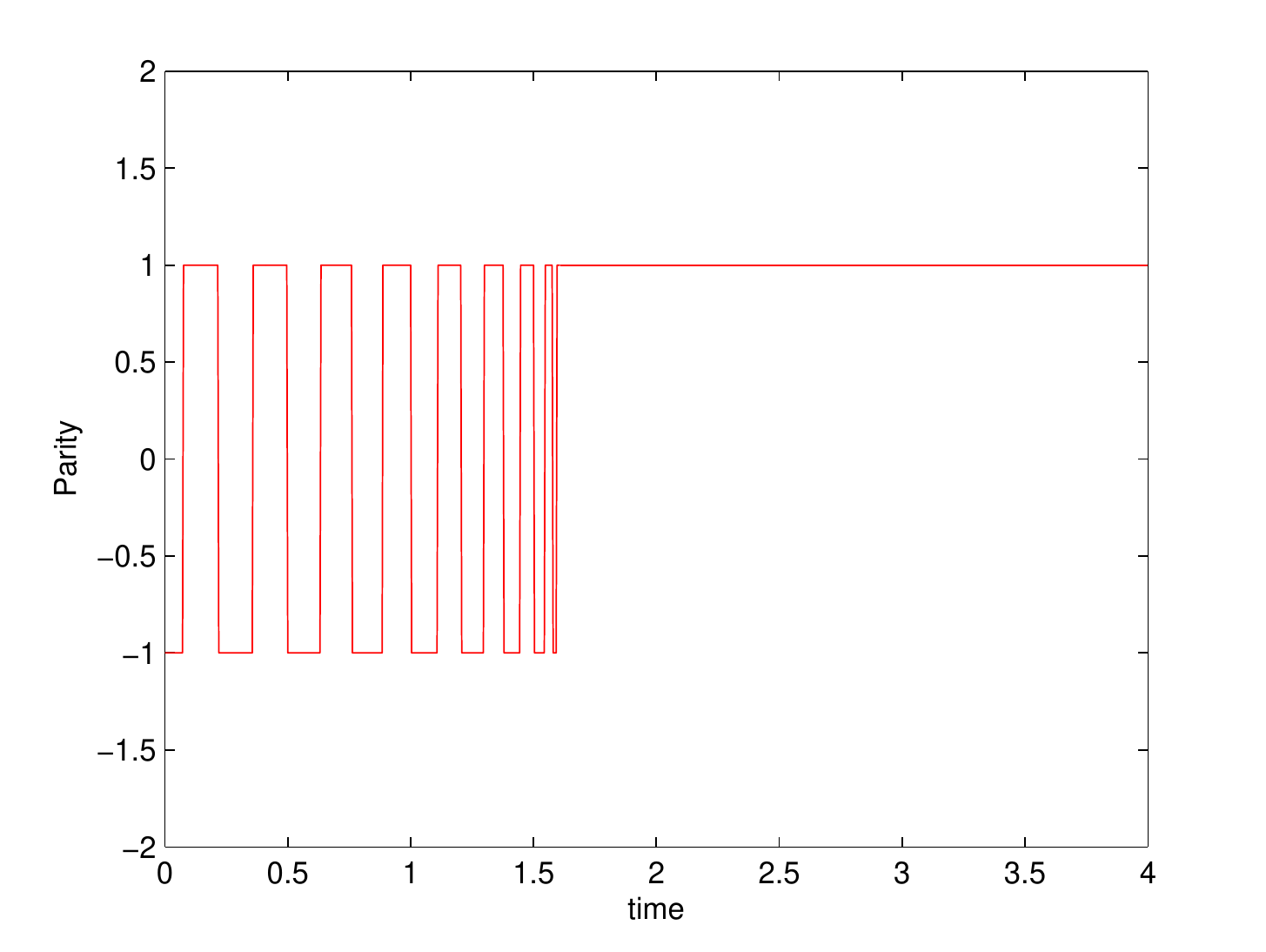}}

\caption{(Color online) Numerical results for the (a) adiabatic fidelity $\mathcal{O}(t)$ and (b) parity of
the instantaneous ground state for an even number of sites. The times at which
the parity switches its sign are exactly the points where parity blocking
occurs, resulting in the adiabatic fidelity plummeting down to zero.
Depending on the parameters chosen, the parity after crossing the quantum
critical point changes from the initial ground state parity thereby leading
to parity blocking for the entire topologically trivial region.}
\label{fig:OverlapOpen34}
\end{figure*}

 In Fig.~\ref{fig:OverlapOpen34}, we present some of our key results. Initializing the system in the ground state corresponding to a specific on-site chemical potential, and thus some fixed parity, we sweep the chemical potential to undergo several parity switches. As a measure of how closely the time evolved state tracks the instantaneous ground state, we evaluate the wave function overlap (adiabatic fidelity) associated with these two states. As seen in  Fig.~\ref{fig:OverlapOpen34}, the adiabatic fidelity plummets down to zero in certain chemical potential intervals that exactly correspond to the parity switched regions. The initial ground state, while being able to track some of the dynamic evolution, is thus forced to remain within its parity sector, an attribute of the topological phase.
This multiple parity blocked dynamics is a dramatic, topologically induced deviation from the continuous evolution expected in quench dynamics.
 
  In what follows, we detail several aspects leading up to this quench behavior, including the formulation of the Majorana wire model, parity switching due to coupling of Majorana wavefunctions and a real space formalism to compute the many-body dynamics. 

\section{$p$-wave superconducting wire}
\label{sec:pwave}
The system that forms our subject of study is a lattice version of the spinless fermionic
 one-dimensional $p$-wave superconducting wire with spinless fermions, also referred
to as the Kitaev chain~\cite{Kitaev01}.
This system can be mapped exactly to a spin-1/2 $XY$ model in a transverse
field (i.e., a magnetic field applied along the $z$ direction) via
 the Jordan-Wigner transformation ~\cite{Lieb61}. The
parameters of the system are the nearest-neighbor hopping amplitude $w$,
the superconducting pairing amplitude between nearest neighbors $\Delta$, and
a chemical potential $\mu$. This model is a paradigm system for demonstrating
numerous interesting topological properties, including the existence of
Majorana modes at the ends of an open chain in the topological phase.

The Hamiltonian of such a system with $N$ sites and open boundary conditions
is given by
\bea H &=& - \sum\limits_{n=1}^{N-1} (-w f^{\dagger}_{n+1}f_n +
\Delta f^{\dagger}_{n+1}f^{\dagger}_n + H.c.) -\mu \sum_{n=1}^{N}(f^{\dagger}_n f_n-1/2), \label{ham1} \eea
where $f_m$ are Dirac fermion operators obeying the commutation
relations $\{f_m,f_n\}=0$ and $\{f_m,f_n^{\dagger}\}=\delta_{mn}$.
We now introduce $2N$ Majorana fermion operators as $a_{2n-1}=
f_n+f_n^{\dagger}$ and $a_{2n}=i(f_n^{\dagger}-f_n)$. These satisfy the
relations $a_m^{\dagger}=a_m$ and $\{a_l,a_m\}= 2 \delta_{lm}$.
In terms of these operators, the Hamiltonian takes the form

\bea \non
H&=&-\frac{i}{2} \sum\limits_{n=1}^{N-1}\bigg[(w-\Delta) a_{2n-1} a_{2n+2}
- (w+\Delta) a_{2n} a_{2n+1} \bigg] -\frac{i\mu }{2} \sum\limits_{n=1}^{N} a_{2n-1} a_{2n}. \label{ham2} \eea

\begin{figure}[hp]
\centering
\includegraphics[width=0.5\textwidth]{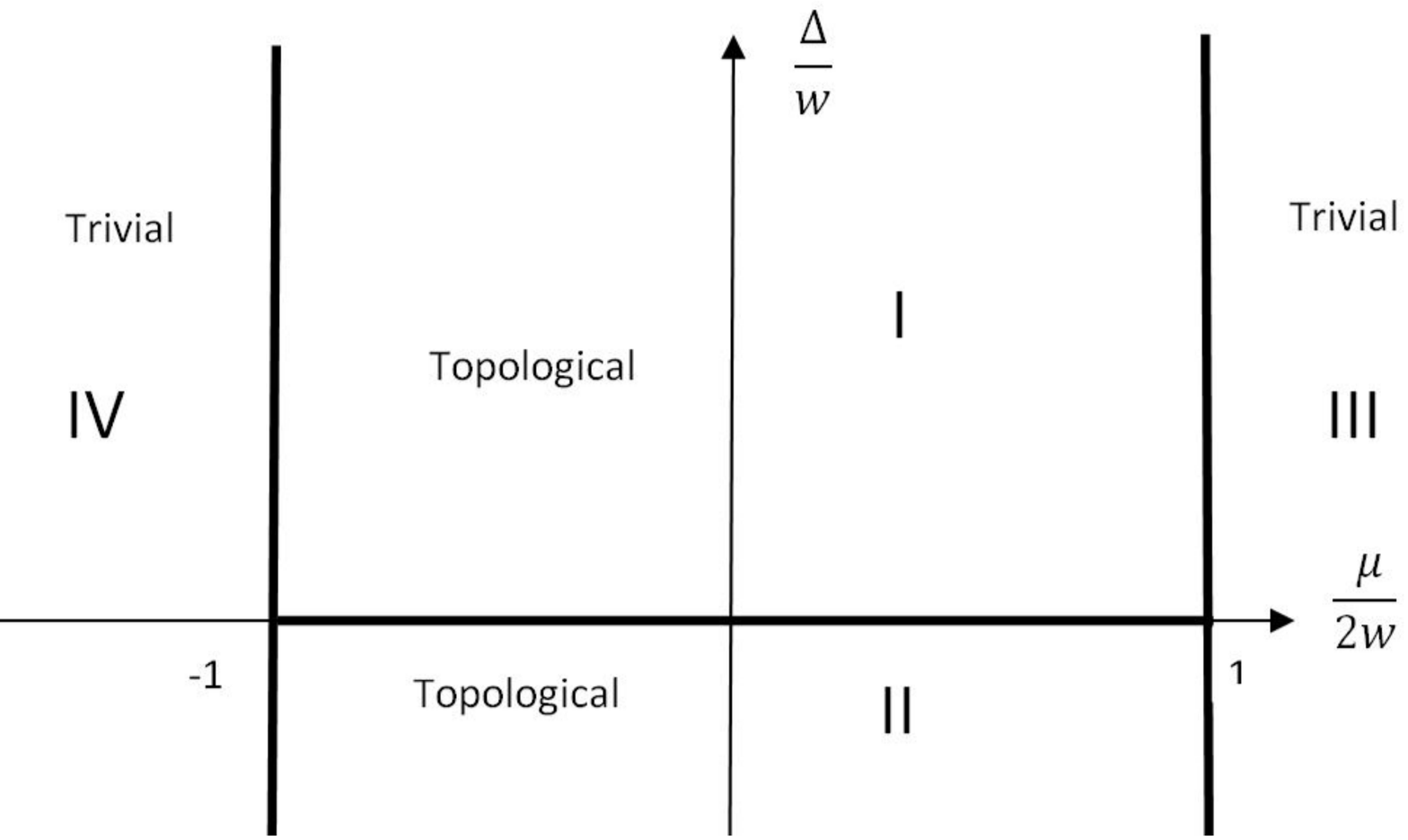}
\caption{The phase diagram of the one-dimensional Kitaev Hamiltonian for the Majorana wire. Phases I and
II are topologically non-trivial and have Majorana end modes, whereas phases
III and IV are topologically trivial. The thick lines $\mu=\pm 2w$ and
$\Delta=0$ are the quantum critical lines where the bulk gap vanishes.}
 \label{fig:phasediagram}
\end{figure}

The Hamiltonian in Eq. (\ref{ham1}) can be diagonalized, up to a constant, to
give the canonical form,
\beq H=4\sum\limits_{j=1}^N \lambda_j b^{\dagger}_j b_j, \eeq
where the $\lambda_j$ are non-negative real numbers. This can be done
through a transformation of the form
\beq \bar{b}=B\bar{a}, \label{bBa} \eeq
where $\bar{a}=(a_1,a_2, \cdots, a_{2N})^T$ and $\bar{b}=(b_1, \cdots,
b_N,b^{\dagger}_1, \cdots, b^{\dagger}_N)^T$ are column vectors with $2N$
components.
The $(2N)$-dimensional matrix $B$ comprises of the eigenvectors of
$H$ and their Hermitian conjugates, belongs to the unitary group
$U(2N)$, and has its determinant satisfying the property $det(B)= \pm 1$.
 The energy eigenvalues of the Hamiltonian are $0$ and $4 \lambda_j$.

The phase diagram of this model is shown in Fig.~\ref{fig:phasediagram}. The
phases I and II are topologically non-trivial and, in the thermodynamic limit,
 have zero energy Majorana
modes bound to the ends of the wire, whereas such modes are absent in the
topologically trivial phases III and IV. These Majorana modes have finite
support at the ends and decay rapidly into the bulk with a decay length
proportional to the reciprocal of the bulk gap. One can understand the
existence of the Majorana end modes by considering the extreme limit of
$w=\Delta$ and $\mu=0$. The Hamiltonian reduces to $H=iw\sum_n a_{2n}
a_{2n+1}$. The Majorana operators $a_1$ and $a_{2N}$ are not paired with
any other operators in the system and therefore do not appear in the
Hamiltonian. These isolated modes correspond to the zero energy eigenvectors
localized at the ends. The existence of these modes is robust even away
from this extreme limit and they only disappear with the closing of the
bulk gap.

The ground state of the system in the topological phase is thus doubly degenerate
and has two zero energy eigenvalues corresponding to the Majorana modes.
These Majorana modes can be combined to form a complex Dirac fermion state,
which can be either empty or occupied. Hence, each of the degenerate ground
states has a specific fermion parity and the system can be characterized by
a related $Z_2$-valued topological invariant. This ground state parity will
play an important role in the subsequent sections.

There are three phase boundaries, indicated by dark lines in
Fig.~\ref{fig:phasediagram}, where
the bulk gap vanishes. 
These are the quantum critical lines across which there is a topological
phase transition. In the thermodynamic limit (infinite wire) or for a
closed chain,
one can transform the Hamiltonian into Fourier space, and the single particle
energy spectrum takes the form
\beq E_k = \pm \sqrt{(2w \cos k+\mu)^2+ 4 \Delta^2 \sin^2k}. \eeq
This spectrum has a finite superconducting gap in all the phases; the gap
vanishes as one crosses one of the critical lines and reopens upon entering
another phase. In the spin language, the topological phases correspond to
the ferromagnetic phases of the transverse field $XY$ model (where either
the $x$ or the $y$ component of the spins has long range order), and the
trivial phases are in the paramagnetic phase.

 These characteristic features of the system, namely, the topological
invariant, the spectrum of the bulk and end modes, and the wave functions of
the Majorana end modes have been discussed extensively in previous work
(see, for example, Ref.~\cite{Kitaev01,DeGottardi11,Alicea12}). So far,
all the above
mentioned characteristics of the model assume the size of the system to be
much larger than the decay length of the Majorana end modes. In the next
section we consider the
case when the Majorana modes at the two ends have a finite overlap,
giving rise to consequences such as parity blocking in quench dynamics.

\section{Finite size effects in the Majorana wire}
\label{sec:finitesize}

\subsection{Tunneling between Majorana end modes}

In a Majorana wire of finite length, the two Majorana end modes are no longer
completely decoupled since there is some overlap between their wave functions
(Fig.~\ref{fig:cartoon}); the overlap shifts their energies slightly away
from zero. In the extreme
limit of the topological phase with $\Delta = w$, the end modes are exactly
localized at the ends with no overlap between them;
hence the effective Hamiltonian governing these modes is
\beq H_f =iJ a_1 a_{2N}, \eeq
with $J=0$, But in general, the effective Hamiltonian has an
expression in terms of the (almost) zero energy eigenvectors
localized at the ends of a finite length wire~\cite{Kitaev01}
\beq H_f=iJ b'b'', \eeq
where $b'=\sum\limits_j(\alpha'_+ x^j_+ +\alpha'_- x^j_-)a_{2j-1}$,
$b''=\sum\limits_j(\alpha''_+ x^{-j}_+ +\alpha''_- x^{-j}_-)a_{2j}$. Here $J$
is in general a function of $\Delta/w$, $\mu/w$ and $N$. Due to this coupling
the two zero modes split in energy into a particle-hole symmetric pair of
eigenenergies $E =\pm J$, and the ground state is no longer
degenerate. The Majorana end modes can be combined into
non-local Dirac fermions as $\tilde{c}=(b'+ib'')/2$ and
$\tilde{c}^{\dagger}=(b'-ib'')/2$. The Hamiltonian can now be expressed as
\beq H_f =J(2\tilde{c}^{\dagger}\tilde{c}-1). \eeq
The occupation number $\tilde{c}^{\dagger}\tilde{c}$ can either be zero or 1.
Thus we see that the energies $\pm E$ come with corresponding eigenstates
with opposite fermion parities. The sign of $J$ decides which of these
states is the ground state. The parity of the states in the bulk being fixed,
the overall ground state parity is then decided by the lower one of the two
split energy levels (which lie inside the bulk gap).
The coupling $J$ is a function of the chemical
potential $\mu$ and oscillates, switching its sign at specific values of $\mu$.
Therefore the split energy levels cross zero at certain points in time
if $\mu$ is varied linearly in time.
This leads to oscillations in the overall parity of the ground state.
Even though these split energy levels are exponentially smaller (for large
system size $N$) than the energies in the bulk, the parity oscillations play
a key role in the time evolution of the ground state in the quenching dynamics.

\begin{figure*}[htp]
\centering
\subfloat[]{\includegraphics[width=0.4\textwidth]{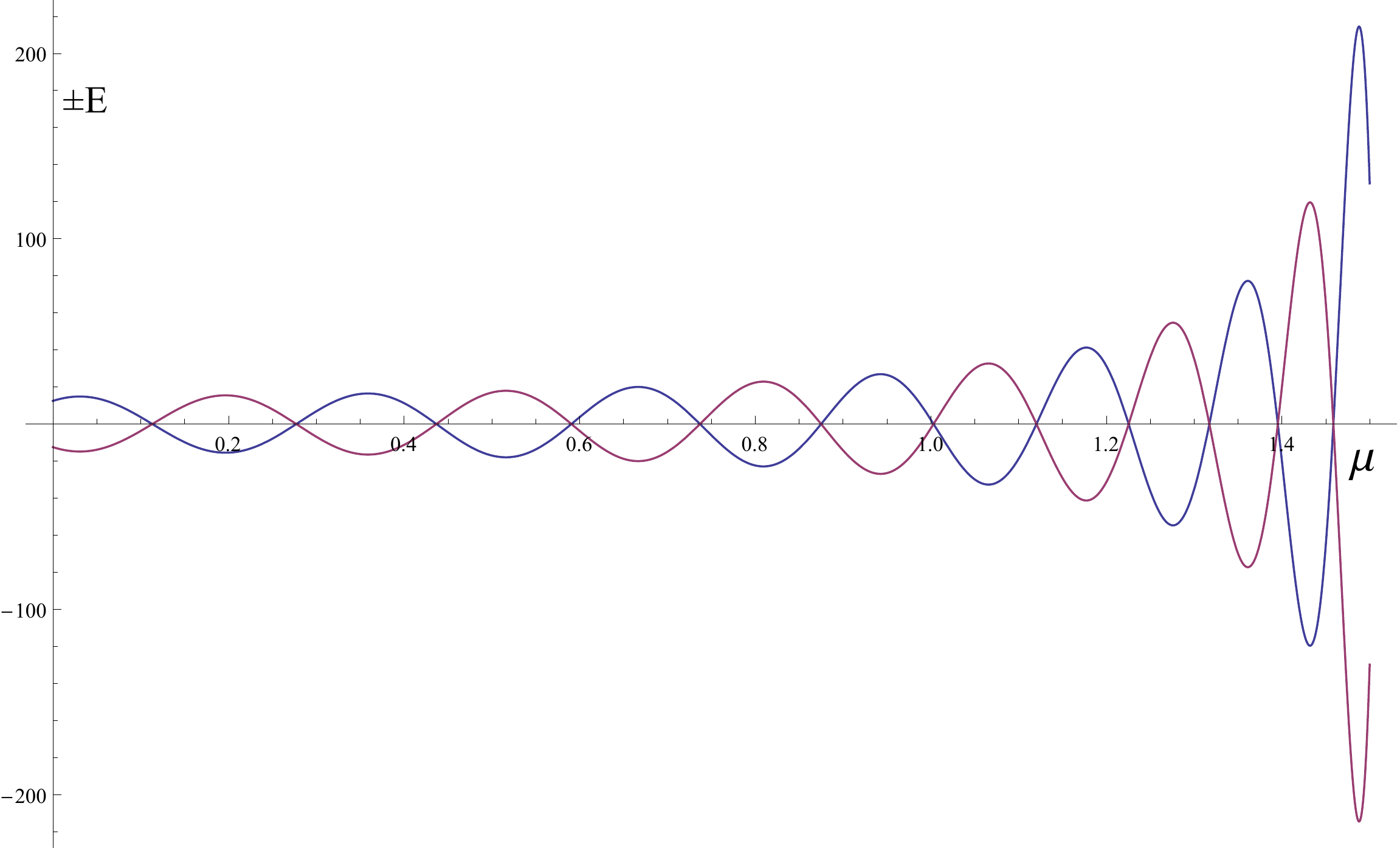}\label{fig:Splittinganalytic}} \hspace{1cm}
\subfloat[]{\includegraphics[width=0.4\textwidth]{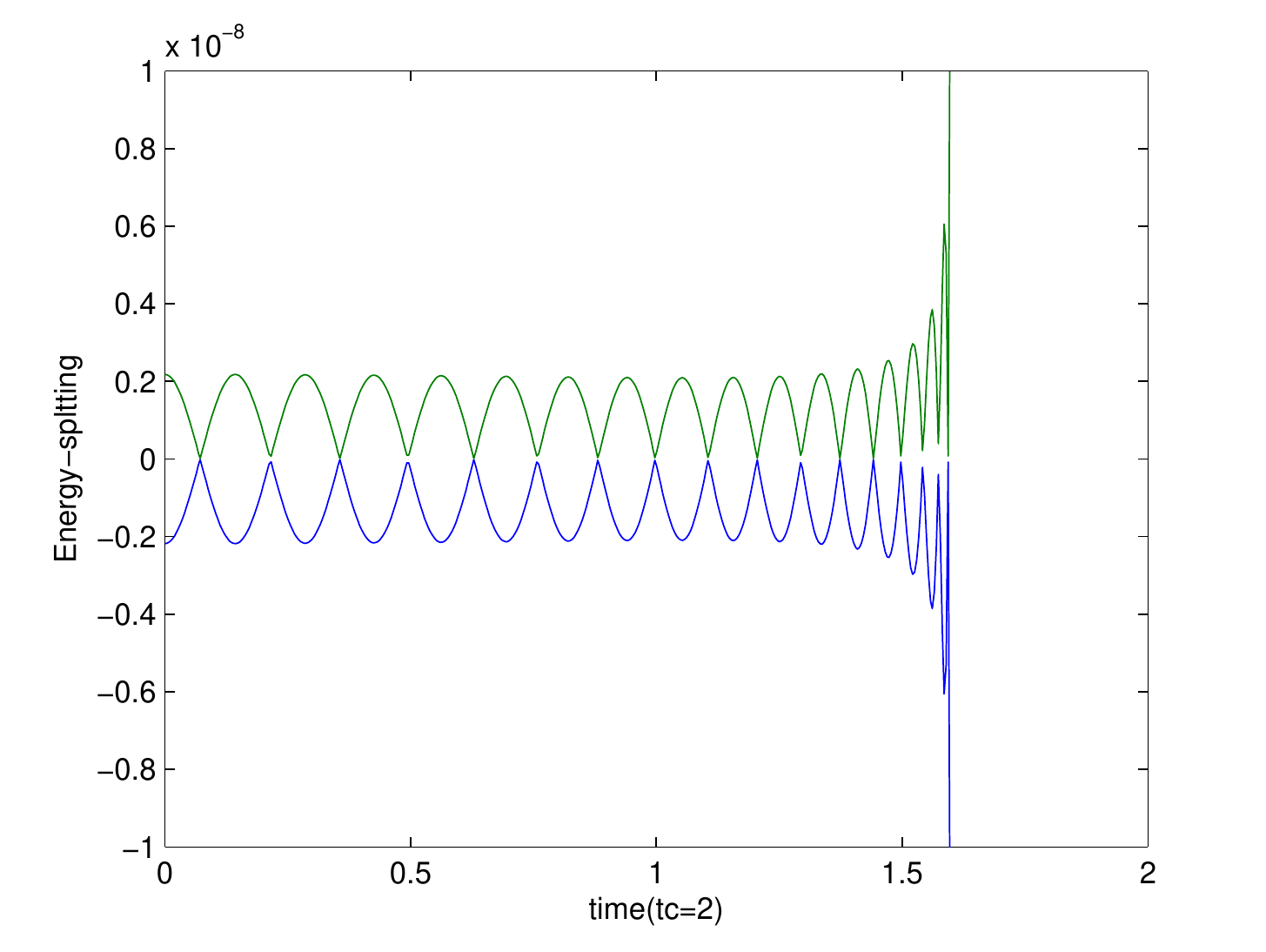}\label{fig:SplittingOpen35}}\\
\subfloat[]{\includegraphics[width=0.4\textwidth]{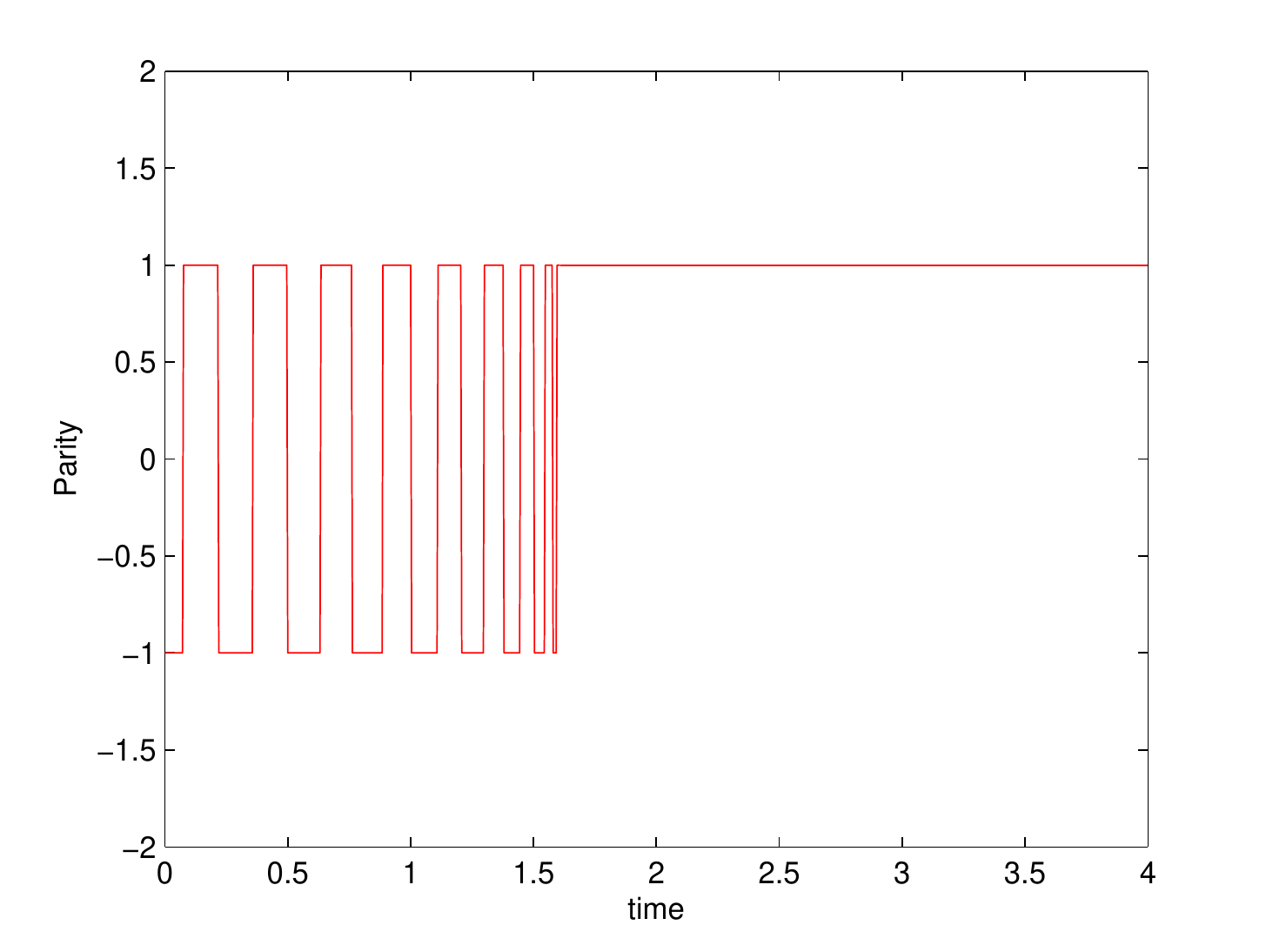}\label{fig:ParityOpen35}}
\caption{(Color online) Oscillations of the split mid-gap energy levels for a linear time variation
of $\mu$. Figure (a) shows the overlap of analytically calculated Majorana wave functions as an indicator of the splitting, and (b) shows the variation of the mid-gap states obtained by numerical diagonalization of the Hamiltonian. Figure (c) shows
that the oscillations of the energy levels in (b) correspond exactly to the
parity oscillations of the ground state obtained numerically from Eq.\ref{detB} in the text.}
\end{figure*}

\begin{figure*}[htp]
\centering
\subfloat[][Energy splitting not crossing zero]{\includegraphics[width=0.4\textwidth]{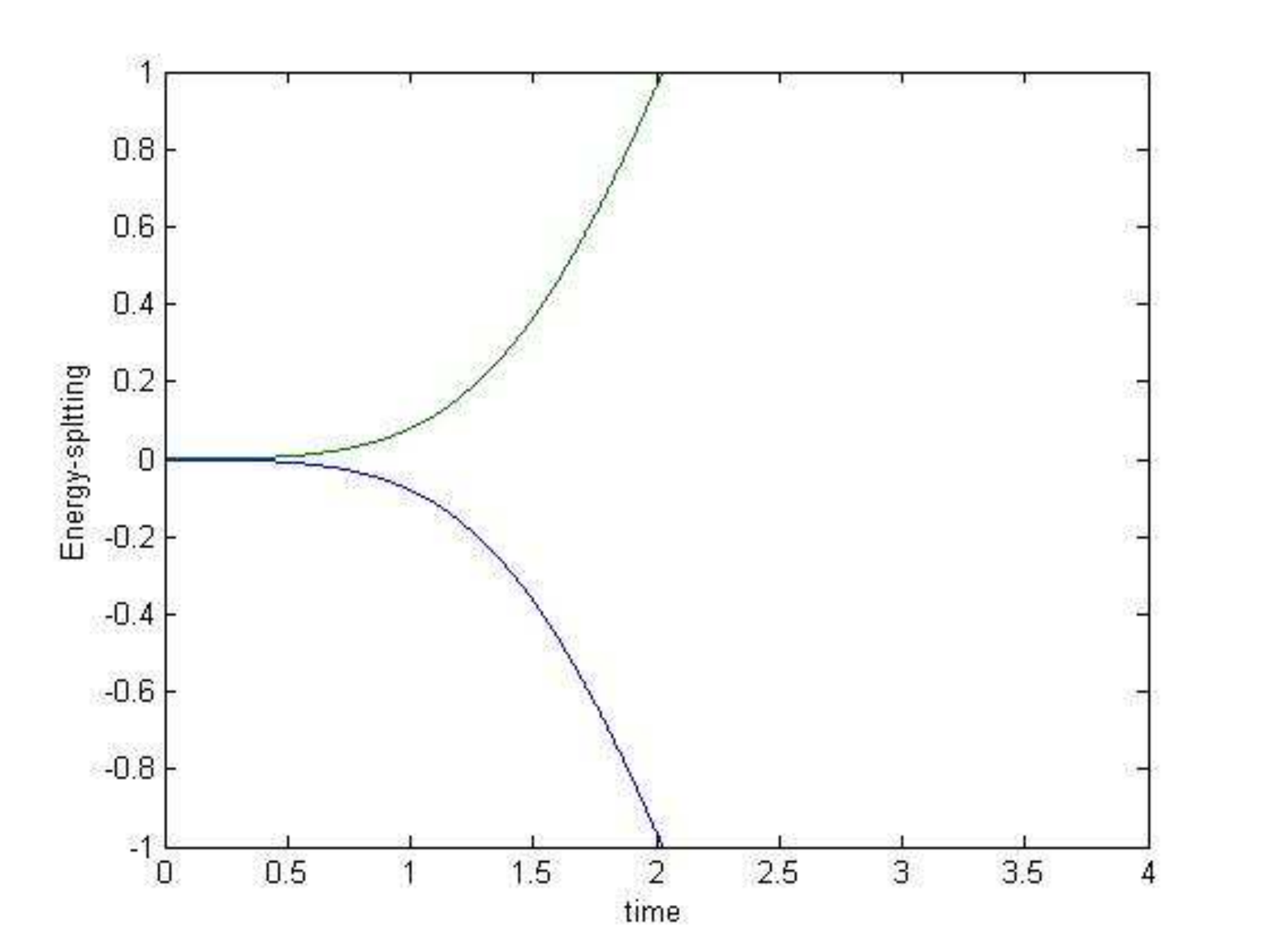}\label{fig:SplittingNoOsc}} \hspace{1cm}
\subfloat[][No parity switching]{\includegraphics[width=0.4\textwidth]{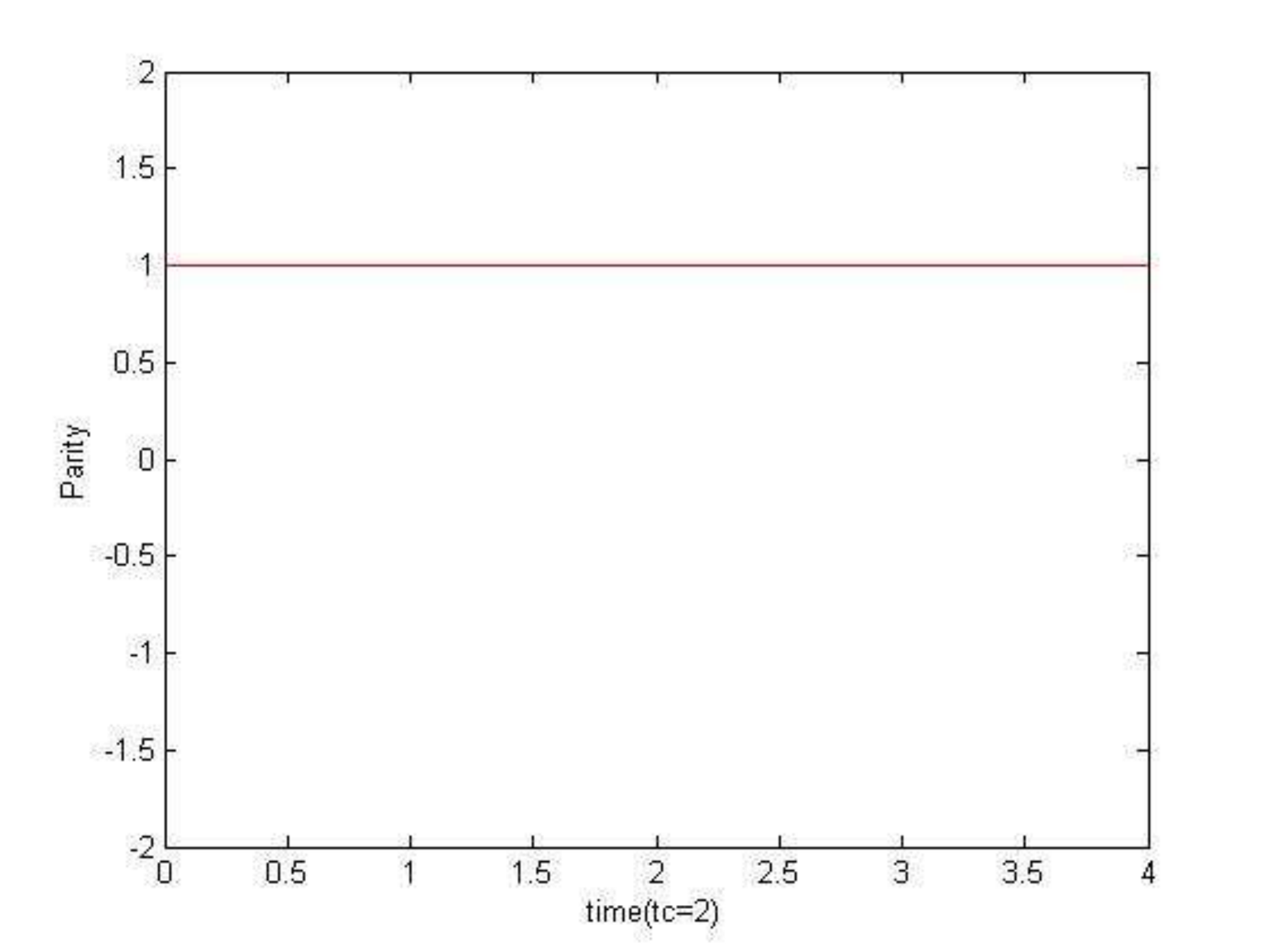}\label{fig:ParityNoOsc}}
\caption{(Color online) Absence of oscillations in the mid-gap levels in the regime which lies outside the circle $\mu=2
\sqrt{w^2-\Delta^2}$. In this regime the parity of the ground state will not show oscillations.}
\end{figure*}

The energy splitting due to the coupling of the two Majorana end modes can be
derived by evaluating the overlap between their associated wave functions.
For a very long wire, the Majorana modes have zero energy; we calculate their
exact wave functions by using the Heisenberg
equations of motion $[H,a_n]=0$ and $[H,b_n]$, where $a_n$, $b_n$ are the
two Majorana operators at site $n$
which were denoted by $a_{2n-1}$ and $a_{2n}$ above. We obtain the difference
equations for these operators as:
\bea (w+\Delta) b_{n-1} + \mu b_n + (w - \Delta)b_{n+1} &=& 0,
\non \\
(w-\Delta) a_{n-1} + \mu a_n + (w + \Delta)a_{n+1} &=& 0, \label{abn1} \eea
for $2 \le n \le N-1$.


These difference equations can be solved exactly by using $z$-transform methods
and taking into account the form of the difference equations at the ends.
The wave functions $\alpha_n$ and $\beta_n$ for the $a_n$ and $b_n$ modes on
site $n$, respectively, are of the form
\bea \alpha_n &=& \alpha_0 C^{n} \bigg[\cos(\omega n)+
\frac{1}{\tan\omega}\sin(\omega n) \bigg], \non \\
\beta_n &=& \beta_0 C^{-n}[\cos(\omega n)+ \frac{1}{\tan\omega}\sin(\omega n)],
\label{alphabetan} \eea
where $C=\left( \frac{w+\Delta}{w-\Delta} \right)^{1/2}$, and
\beq \omega= \tan^{-1} \left( \frac{\sqrt{4w^2-4\Delta^2-\mu^2}}{
\mu} \right). \label{om} \eeq
Using these wave functions, we can now approximately calculate the energy
splitting by assuming that the wire has a finite length $N$
and computing the overlap
of the wave functions $\alpha_L$ at the left end and $\alpha_R$ at the
right end. This gives an expression of the form
\bea \non
\alpha_L \alpha_R &=&\frac{\alpha_{0L} \alpha_{0R} a^{N+1}}{4\sin^3\omega}
\bigg[2\sin(\omega N)\\ &&+ N\{\sin[(N+2)\omega] - \sin[(N+4)\omega]\}\bigg].
\label{aLR} \eea

Figure~\ref{fig:Splittinganalytic} shows the oscillations in the splitting as calculated above.
This is an approximate calculation because we have assumed the energies to
be zero in the Heisenberg equations of motion for the Majorana operators and
then calculated their overlap (which shifts the energies slightly away from
zero). But one can see that it qualitatively agrees
with the exact numerical calculation in Fig.~\ref{fig:SplittingOpen35}. Figure~\ref{fig:ParityOpen35} shows
the oscillations of the ground state parity as a function of $\mu$ which is
linearly varied with time. One sees an excellent correspondence with the
zero crossings of the energy splitting in Fig.~\ref{fig:SplittingOpen35}. Further, although
these oscillations appear due to the degeneracy splitting in the
topological phase, they do not exist in the entire topological region in the
phase diagram. To understand this, we stress the fact that the key ingredient
in getting these oscillations is the oscillatory component in the wave
functions of the Majorana modes given in Eqs. (\ref{alphabetan}). One can see
that the oscillatory functions $\sin(\omega n)$ and $\cos(\omega n)$ become
hyperbolic if $\omega$ given by Eq. (\ref{om})
becomes imaginary. The boundary at which this happens is given by the
circle $\mu^2=4(w^2-\Delta^2)$. Beyond this circle the Majorana wave functions
have only a decaying (but not oscillatory) component. This implies that
there would not be any parity oscillations in this region. This can be
clearly seen in Fig.~\ref{fig:SplittingNoOsc}.

We remark here that expressions for the energy splitting
of the Majorana end modes are known for a continuum
model~\cite{Pientka13,Thakurathi14}. These results are consistent with the 
 splitting being both oscillatory
and decaying exponentially with increasing length.

{\bf Calculation of ground state fermion parity:}
To obtain a rigorous knowledge of the parity and its switching as a function
of the chemical potential, as has been used in other Majorana wire
contexts~\cite{Sau13},
we employ the measure introduced in Kitaev's well-known
work~\cite{Kitaev01}. Given a
Hamiltonian of the form in Eq. (\ref{ham1}), the transformation $B$, which
reduces the Hamiltonian to the canonical form, can be represented as a
conjugation by a parity preserving unitary operator if $B$ has the form
$B=e^D$ i.e., if $det(B)=1$. Otherwise $B$ changes the parity.
Therefore, the parity of the system is given by
\beq P(H)=sgn[det(B)]. \label{detB} \eeq
In Appendix B we illustrate this result with a simple problem of a
two-site effective Hamiltonian. This illustration is particularly useful in our
case since the effective Hamiltonian for only the coupled
Majorana modes is in fact a two-site problem given by $H_f=iJb' b''$.
Within the topological phase, the dynamics of only these end modes and their
associated Dirac fermions determine the overall parity as we saw above.

In terms of the Majorana operators, the parity of a $N$-site system is
given by
\beq P~=~ \prod_{j=1}^N ~(-ia_{2j-1}a_2j) ~=~ \prod_{j=1}^N
~(1 - 2 f_j^\dagger f_j). \label{parity} \eeq
We note that $P$ is both Hermitian and unitary, and it commutes with the 
Hamiltonian in Eq. (\ref{ham2}); since $P^2 = I$, the eigenvalues of $P$ must 
be $\pm 1$.
In the extreme case of $w=\Delta$ for the Majorana wire in Eq. (\ref{ham2}),
the terms of the form $a_{2j}a_{2j+1}$ with $j=1, \cdots, N-1$ are equal to 1.
Thus only the term $a_1a_{2N}$ remains, which is in fact the term in the
effective Hamiltonian. Now, within the topological phase, small deformations
of the parameters should not change this fact. Since the parity operator also
commutes with the Hamiltonian, we can see that in the topological phase
the end modes alone determine the parity of the ground state. The parity equivalence of ground state sectors has been studied in ~\cite{Kells14c}

\subsection{Exact expression for zero energy contours}

While we obtained an approximate result for the tunneling amplitude that
splits the degeneracy between Majorana end modes in the previous subsection,
a formalism involving transfer matrices~\cite{DeGottardi11,DeGottardi13}
enables to track the exact points in the parameter space at which this
energy changes
sign, restoring the zero energy degeneracy and resulting in a parity switch.
Previous work has presented similar derivations and results using the
equivalent method of chiral decomposition~\cite{Kao14}.

For a mode with energy exactly equal to zero, Eqs. (\ref{abn1}-\ref{alphabetan})
are applicable. We see that the $a$ and $b$ modes are decoupled; for
definiteness, let us consider a zero energy mode involving the $a_n$'s.
Eq. (\ref{abn1}) shows that for $2 \le n \le N-1$, the $a_n$'s are related to
each other by a transfer matrix $M_a$,
\bea \left( \begin{array}{c}
a_{n+1} \\
a_n \end{array} \right) &=&~ M_a \left( \begin{array}{c}
a_n \\
a_{n-1} \end{array} \right), \non \\
\mbox{where}~~ M_a &=& \left( \begin{array}{cc}
-\frac{\mu}{w + \Delta} & - \frac{w - \Delta}{w + \Delta} \\
1 & 0 \end{array} \right). \eea
If $\lambda_1$ and $\lambda_2$ are the two eigenvalues of $M_a$, the general
solution of Eq. (\ref{abn1}) is $a_n = c_1 \lambda_1^n + c_2 \lambda_2^n$.
Next we note that the boundary equations become satisfied
if we add fictitious sites with $n=0$ and $n=N+1$ at the two ends of the
system and demand that $a_0 = a_{N+1} = 0$. This is possible if and only if
\beq \left( \frac{\lambda_1}{\lambda_2} \right)^{N+1} ~=~ 1. \label{la12} \eeq

We can show that Eq. (\ref{la12}) holds and that hence there is a zero energy
mode if either \\
\noindent (i) the parameters lie insider the circular region $\mu^2 + 4
\Delta^2 < 4 w^2$, and
\beq \left( \frac{\mu + i \sqrt{4 w^2 - 4 \Delta^2 - \mu^2}}{\mu - i
\sqrt{4 w^2 - 4 \Delta^2 - \mu^2}} \right)^{N+1} ~=~ 1, \label{zeroen} \eeq
or \\
\noindent (ii) $\mu = 0$ (which implies that $\lambda_1 /\lambda_2 = -1$
regardless of the relative values of $w$ and $\Delta$) and $N$ is odd. \\
We get the same conditions if we look for a zero energy mode involving the
$b_n$'s.

In terms of $\omega$ defined in Eq. (\ref{om}), Eq. (\ref{zeroen}) is
equivalent to saying that $(N+1)\omega$ is an integer multiple of $\pi$,
namely, that $\sin [(N+1)\omega] = 0$. We see that this differs somewhat
from the approximate condition that $\alpha_L \alpha_R$ given in
Eq. (\ref{aLR}) should be equal to zero.

\begin{figure*}[htp]
\centering
\subfloat[][Parity variations for even number of sites (N=10).]{\includegraphics[width=4in]{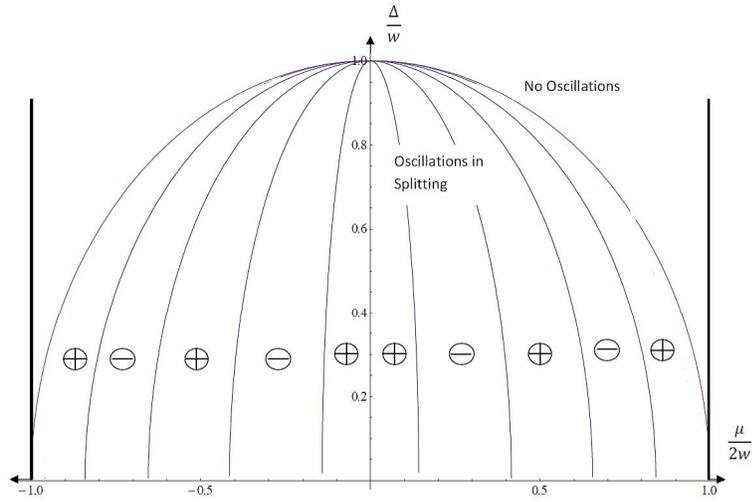}} \hfill
\subfloat[][Parity variations for odd number of sites (N=11).]{\includegraphics[width=4in]{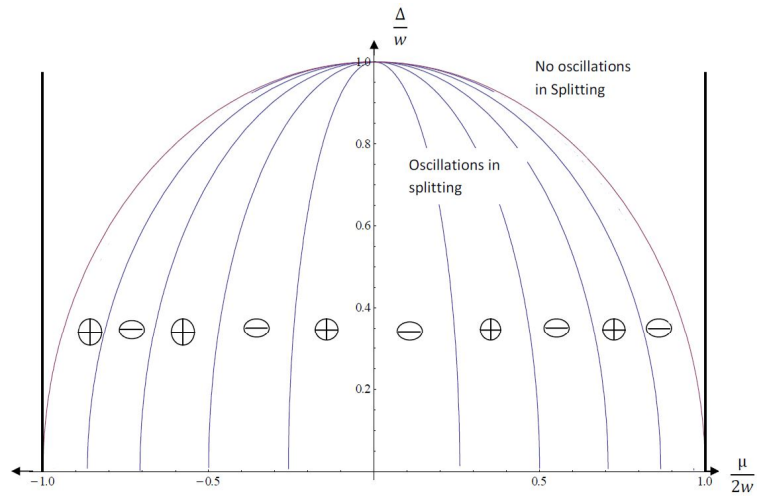}}
\caption{(Color online) The phase diagram of the Kitaev Hamiltonian for the Majorana wire representing regions of odd and
even fermion parity ($\pm$) for finite length wires. The outer circle is the boundary at
which the oscillations in the Majorana wave function and the energy splitting
stops. Contours corresponding to exact zero energy states and parity switching
form ellipses. The parity corresponding to the specific parameters changes for
odd and even number of sites. As we can see there is a contrast in the behavior of parity for even and odd $N$ across the $\mu=0$ line. For odd $N$, the
parity actually changes across this line and thus starting from $\mu=0$ gives a special case of parity blocking. }\label{fig:oddNparitydiagram}
\end{figure*}

The solutions of Eq. (\ref{zeroen}) are given by
\beq 4 \Delta^2 + \mu^2 \sec^2 \left( \frac{\pi p}{N+1} \right) ~=~ 4 w^2,
\label{ellipse} \eeq
where $p$ is an integer equal to $1,2,\cdots,N/2$ if $N$ is even and $1,2,
\cdots,(N-1)/2$ if $N$ is odd. In terms of the variables $\mu/w$ and
$\Delta/w$, we observe that Eq. (\ref{ellipse}) defines a number of ellipses,
which are labeled by the integer $p$; these are shown in
Fig.~\ref{fig:oddNparitydiagram} (a) for $N$
even and Fig.~\ref{fig:oddNparitydiagram} (b) for $N$ odd. Note that all the ellipses pass through the two
points given by $\mu = 0$ and $\Delta = \pm w$.
Fig.~\ref{fig:oddNparitydiagram} (b) for $N$ odd also
contains a zero energy line lying at $\mu = 0$ for all values of $\Delta/w$.

Eq. (\ref{ellipse}) can be understood in a simple way for the special case
$\Delta = 0$. Eq. (\ref{ellipse}) then reduces to
\beq \mu ~=~ \pm ~2 w ~\cos \left( \frac{\pi p}{N+1} \right). \label{Delta0}
\eeq
We can understand this as follows. For $\Delta = 0$, Eq. (\ref{ham1})
describes a non-superconducting tight-binding model whose single particle
spectrum is given, for an open chain with $N$ sites, by $E_q = 2 w \cos
[\pi q/ (N+1)] - \mu$, where $q = 1, 2, \cdots, N$. One of these energies
vanishes whenever $\mu$ satisfies the condition given in Eq. (\ref{Delta0}),
in particular, when $p$ is equal to the smaller of the two integers $q$ and
$N+1-q$. This is where the ranges of $p$ mentioned above, namely, $p=1,2,
\cdots,N/2$ for $N$ even and $1,2,\cdots,(N-1)/2$ for $N$ odd come from.
In addition, if $N$ is odd (but not if $N$ even), we have a zero energy
state at $\mu = 0$ corresponding to $q=(N+1)/2$.

Having found all the zero energy lines in the plane defined by $\mu/w$ and
$\Delta/w$, we observe that the parity of the fermion number of the ground
state flips sign whenever we cross one of these lines. As a check, this is
again easy to
see for the case $\Delta = 0$. The number of energy levels which are occupied
in the ground state changes by 1 and hence the fermion parity
changes sign whenever one of the single particle energies $E_q$ given above
crosses zero. For $\Delta = 0$ and very large negative values of $\mu$,
we can see that the ground state of Eq. (\ref{ham1}) contains no fermions;
hence, according to Eq. (\ref{parity}), the fermion parity is $+1$ for any
value of $N$. For very large positive values of $\mu$, the ground state of
Eq. (\ref{ham1}) is completely filled with $N$ fermions; hence the
fermion parity according to Eq. (\ref{parity}) is $(-1)^N$.
We remark that the oscillations in the parity, which are related to the Kitaev's
Pfaffian, map to spin-spin correlations in the transverse spin chain, and as with
much of the literature on Majorana wires, these oscillations have been discussed
in depth in the spin context~\cite{Barouch}.

\section{Quenching dynamics in the Majorana wire}
\label{sec:quenching}
Previous work involving the dynamics of quenching in the Majorana wire
described above has focused on tuning through quantum critical points
separating topological and trivial phases~\cite{DeGottardi11,Kells14a,Bermudez10,Rajak14}.
There have been recent works on the effect of quenching on Majorana modes, signatures of Majorana modes in quenching dynamics and Kibble-Zurek scaling ~\cite{Vasseur14,Perfetto13,MLee14,Karzig14,sacramento14}.
(As mentioned earlier, while the terms quench and ramp
are sometimes used to distinguish between instantaneous change of parameter
versus
a time-dependent change at some given rate, respectively, here we use quench
in a more general sense to encompass all such dynamic tuning. Our actual studies
 are restricted to the ramp case.)
While our analysis also explores non-equilibrium dynamics within a
particular topological phase, we use similar protocols for changing
parameters of the Hamiltonian to tune from one phase to another.

Specifically, we consider a linear variation with time of the chemical
potential of the system so as to go across the critical line $\mu = 2w$,
\beq \mu(t)= (2-\mu_i)t/\tau +\mu_i. \label{mut} \eeq
Here $\mu_i$ is the initial chemical potential at $t=0$ and $1/\tau$ is
the quench rate. Due to the finite rate of variation of $\mu$, the system
cannot remain exactly in its ground state and will exhibit
non-equilibrium behavior. Namely, excitations (or defects) will be produced
in the ground state; this lead to an excess energy of the system and may
also lead to a ground state which is in a different topological sector
than the initial ground state. These effects can be characterized by
the following quantities.
\begin{itemize}
\item Defect density: The number of defects produced in the ground state
configuration, which is given by the sum over all the excitations.
\item Adiabatic Fidelity $\mathcal{O}(t)$: This is the inner product of the
instantaneous ground state $\ket{\psi_{ins}(t)}$ of the time-dependent
Hamiltonian with the time evolved initial ground state $\ket{\Psi(t)}$,
\beq \mathcal{O}(t)=|\langle \Psi(t)\ket{\psi_{ins}(t)}|.
\label{eq:overlap} \eeq

\item Residual energy $E_{res}$: This is the energy in excess of the
instantaneous ground state of the system. We will define this as the
dimensionless quantity
\beq E_{res} = [\langle \Psi(t)| H(t)| \Psi(t)\rangle - E_G(t)]/|E_G(t)|,
\label{eq:eres} \eeq
 where $E_G$ is the energy of the ground state at time $t$.
\end{itemize}

{\bf Previous work:} Most of the analytical results for the above quantities
obtained in earlier work are in the limit of very large system size or
with periodic boundary conditions, where one can Fourier transform the
Hamiltonian to momentum space. This in fact reduces the calculation to a well
known problem of a transition between two states for each value of the
momentum $k$ in the Brillouin zone. This is the famous Landau-Zener-Majorana-Stueckelberg
problem~\cite{Landau32,Zener32,Majorana32,Stueckelberg32} which can,
under a few assumptions, be solved exactly to obtain the probability of
excitation from the ground state to the excited state. Using this probability,
we can obtain expressions for the defect density, adiabatic fidelity and residual energy.
In the limit of long time $t$, all these quantities
have a universal power law scaling as a function of the quench rate which
is related to the post-quench excitations. This is the well studied
Kibble-Zurek scaling~\cite{Dziarmaga10,Polkovnikov11,Dutta10a,Kibble76,
Kibble80,Zurek85,Zurek96,Dziarmaga05,Damski05,Damski06,Polkovnikov05,
Polkovnikov08,Calabrese05,Calabrese06,Cherng06,Mukherjee07,Divakaran08,Deng08,
Divakaran09,Mukherjee10,Sengupta08,Mondal08,Sen08,Mondal09,DeGrandi08,
Barankov08,DeGrandi10,Patane08,Patane09,Bermudez09,Bermudez10,Perk09,Sen10,
DeGottardi11,Pollmann10,Dutta10b,Hikichi10,Chandran12,Chandran13,Patel13,
Mostame14,Foster14}.
Given the quench rate $1/\tau$, the defect density and residual energy scale
as $1/\sqrt{\tau}$ and the adiabatic fidelity scales as $\exp (c/\sqrt{\tau})$
in this one-dimensional Majorana wire.

For the Majorana wire with open boundary conditions, there have been
investigations of the behavior of single particle states under a quench.
While it has been found that the single particle bulk states still obey the
Kibble-Zurek scaling for the defect density, the quench for an initial state
with a Majorana end mode has been found to be non-universal and dependent on
the topological
features of the system~\cite{Bermudez10}. The end states are not robust
with respect to the quench and they delocalize to merge with the bulk states.
This leads to a scaling of the defect density as $\tau^0$ (i.e., independent
of the quenching rate), which is very different from the Kibble-Zurek scaling.

Another investigation for the open chain looks into a quantity called the
Loschmidt echo, which is the survival probability of the Majorana end modes
under a quench~\cite{Rajak14,Vasseur14}. Upon quenching across the critical point, the
probability decays to extremely small values as the end modes merge with the
bulk states when the system is near the critical point where the gap between
the end and bulk states vanishes. But interestingly, when the system is
quenched to exactly the critical point, although the probability of survival
goes to zero initially, it revives itself completely at regular intervals of
time. This is attributed to the nearly equal spacing of the low-lying energy
levels in the bulk near the critical point; this spacing is of the order
of $1/N$ while the scaling of the gap at the critical point is also of order
$1/N$. This leads to oscillations in the survival probability with a time
period which is proportional to the system size $N$. We will see below the
remnant of this effect in the residual energy for an extremely slow quench.

In this work we will study the quenching dynamics in the above mentioned
quantities for an open Majorana wire. We will mainly focus on the many-body
states rather than the single particle states and the novel parity switching
mechanism discussed in the previous section, which comes about in the
topological phase due to the coupling between the Majorana end modes.

\section{Real space formalism for studying quenching dynamics for
open boundary conditions}
\label{sec:realspace}

 In comparison to the translationally invariant systems with periodic boundary conditions studied in previous work, a challenge encountered in these finite-sized systems with open boundary conditions is that one cannot exploit the momentum basis, which in previous works highly simplified the quench dynamical problem. Here,in principle, we are faced with the full-fledged $2^N$ -dimensional Hilbert space associated with fermions on a $N$-site lattice associated with the Fock space formed by fermion occupancy on each site.  Here, we develop and present a dynamic many-body technique to reduce the problem to a numerically tractable form. Our technique hinges on two principles in calculating expectation values or overlaps between states in this time-dependent setting.  The first is to use the Heisenberg picture so that the crux of the information on the time evolution is given by the relation between fermionic creation/annihilation operators at different times. The second is to invoke an analog of Wick’s theorem for Majorana operators. The computation then reduces to dealing with time-dependent $2N \times 2N$ matrices, allowing us to embark on an exhaustive analysis of adiabatic fidelity and residual energies and to pinpoint attributes of parity blocked dynamics.

  Let us start with a general time-dependent Hamiltonian which is
quadratic in Majorana operators $a_j$ ($j=1,2,\cdots,2N$),
\beq H=i\sum\limits_{i,j=1}^{2N} a_i M_{ij}(t) a_j. \eeq
Here $M(t)$ is a real antisymmetric matrix; its elements will be functions
of $w$, $\Delta$ and $\mu$ for the Majorana wire Hamiltonian in Eq.
(\ref{ham2}). This can be converted to the canonical form
\beq H=4\sum\limits_{j=1}^N \lambda_j(t) b^{\dagger}_j(t)b_j(t), \eeq
up to a constant, by a time-dependent transformation $B(t)$,
\beq \bar{b}(t)=B(t)\bar{a}. \label{matB} \eeq
Here $\bar{b}(t)$ is a $(2N)$-component vector $\bar{b}=(b_1,b_2,\cdots,
b_N,b^{\dagger}_1,\cdots,b^{\dagger}_N)^T$ and so is $\bar{a}=(a_1,a_2,
\cdots,a_{2N})^T$. The $(2N)$-dimensional matrix $B(t)$ comprises of the
eigenvectors of $H(t)$ and it belongs to the group $U(2N)$ with $det(B)=
\pm 1$. The eigenvalues of the Hamiltonian are $\pm \lambda_j$.

{\bf Adiabatic fidelity calculation:} As defined in a previous section the
adiabatic fidelity is given by $\mathcal{O}(t)= |\langle \psi_{ins}(t)\ket{\Psi(t)}|$. The corresponding annihilation operators of these many-body states
satisfy the relations
$b_j(t)\ket{\psi_{ins}(t)}=0$ and $\beta_j(t)\ket{\Psi(t)}=0$, where
$\ket{\Psi(t)}=S(t,0)\Psi(0)$ and $\bar{\beta}(t)=B(0)S(t,0)\bar{a}$. Here
$S(t,0)=\mathcal{T}exp(-4 \int_{0}^t M(t')dt')$ is the evolution operator,
with $\mathcal{T}$ denoting time ordering. The two sets of annihilation
operators are related by
\beq \bar{\beta}(t)=B(0)S(t,0)[B(t)]^{-1}\bar{b}(t)=G(t)\bar{b}(t). \label{matS}\eeq
The key idea underlying the calculation in real space is to express the
quantities of interest to us in terms of objects which can be calculated
numerically in a simple way. Given the form of the initial Hamiltonian $H(0)$
and the time-dependent $H(t)$, we can see that the quantities $B(0)$, $B(t)$,
$S(t,0)$ and
$G(t)$ can be easily computed. Given these and the annihilation operators for
the ground states, the calculation of the adiabatic fidelity reduces to a computation of
the determinant of an antisymmetric matrix $A$ given by
\begin{eqnarray*}
A_{jk} &=& \bra{\psi_{ins}(t)}\bar{\beta}_j(t) \bar{\beta}_k(t)
\ket{\psi_{ins}(t)} ~~{\rm for}~~ j<k, \\
&=& -\bra{\psi_{ins}(t)}\bar{\beta}_k(t) \bar{\beta}_j(t)\ket{\psi_{ins}(t)}
~~{\rm for}~~ j>k, \\
&=& 0 ~~{\rm for}~~ j=k. \end{eqnarray*}
We now directly state an important result, deferring the detailed derivation
to Appendix A. The adiabatic fidelity defined in Eq. (\ref{eq:overlap}) is given by
\beq \mathcal{O} (t) = |det(A)|^{1/4}. \label{detlap}\eeq
Given this relation and the Hamiltonian $H(t)$, we can numerically calculate
the adiabatic fidelity as a function of time for a system with open boundary conditions.
This can naturally be extended to periodic/antiperiodic boundary conditions as well.

{\bf Residual energy calculation:}
Another quantity of interest, the residual energy, defined in Eq. (\ref{eq:eres}),
measures the excess energy contained in the time-evolved quench dependent state
compared to the instantaneous ground state energy. This quantity can also
be calculated with the real space formalism developed in this section.
Following the same strategy as for the adiabatic fidelity, the final expression can
simply be expressed in terms of the matrix $G(t)$,
\beq E_{res}=[4 \sum\limits_{j,k}^N \lambda_j G_{N+j,k}^{-1}(t)
G_{j,k+N}^{-1}(t)]/|E_G(t)|. \label{resen} \eeq
The details of the derivation are given in Appendix C.

\section{Results}
\label{sec:results}
We now present the results that we obtain for an open Majorana wire with
$N$ sites with the Hamiltonian given in Eq. (\ref{ham2}), where $\mu$ varies
linearly in time as shown in Eq. (\ref{mut}). Given this specific form of
$H(t)$, we numerically calculate all the quantities $B(0)$, $B(t)$ in
Eq. (\ref{matB}),
$S(t,0)$, $G(t)$ in Eq. (\ref{matS}) and finally the adiabatic fidelity
$\mathcal{O}(t)$ in Eq. (\ref{detlap}) and the residual
energy $E_{Res}$ in Eq. (\ref{resen}). In what follows, we provide
an in-depth discussion of the novel phenomenon of topological and associated
parity blocking as elucidated in Sec.~\ref{sec:blocking}.

For periodic boundary conditions, the effect of the number of sites on
the fermion parity of the ground state and the consequent phenomenon called
topological blocking on the quenching dynamics of the ground state has
been discussed in detail in Ref.~\cite{Kells14a}. For an open chain,
we saw in Sec.~\ref{sec:finitesize} that the Majorana end modes play an
important role in
determining the parity. For a fixed $\Delta$ and $N$, the parity changes
sign as we sweep across the topological phase by varying $\mu$. Here we
will explicitly see this parity blocking effect in the evolution of the
ground state within the topological phase. The initial parity of the
system is set by the value of $\mu_i$ and the number of sites $N$. As
we will see below, the choice of $\mu_i$ can have drastic consequences,
especially for an odd number of sites.

\subsection{Adiabatic fidelity and Parity blocking}

\begin{figure*}[htp]
\centering
\subfloat[][Adiabatic fidelity for $N=17$ with $\mu_i \neq 0$.]{\includegraphics[width=0.4\linewidth]{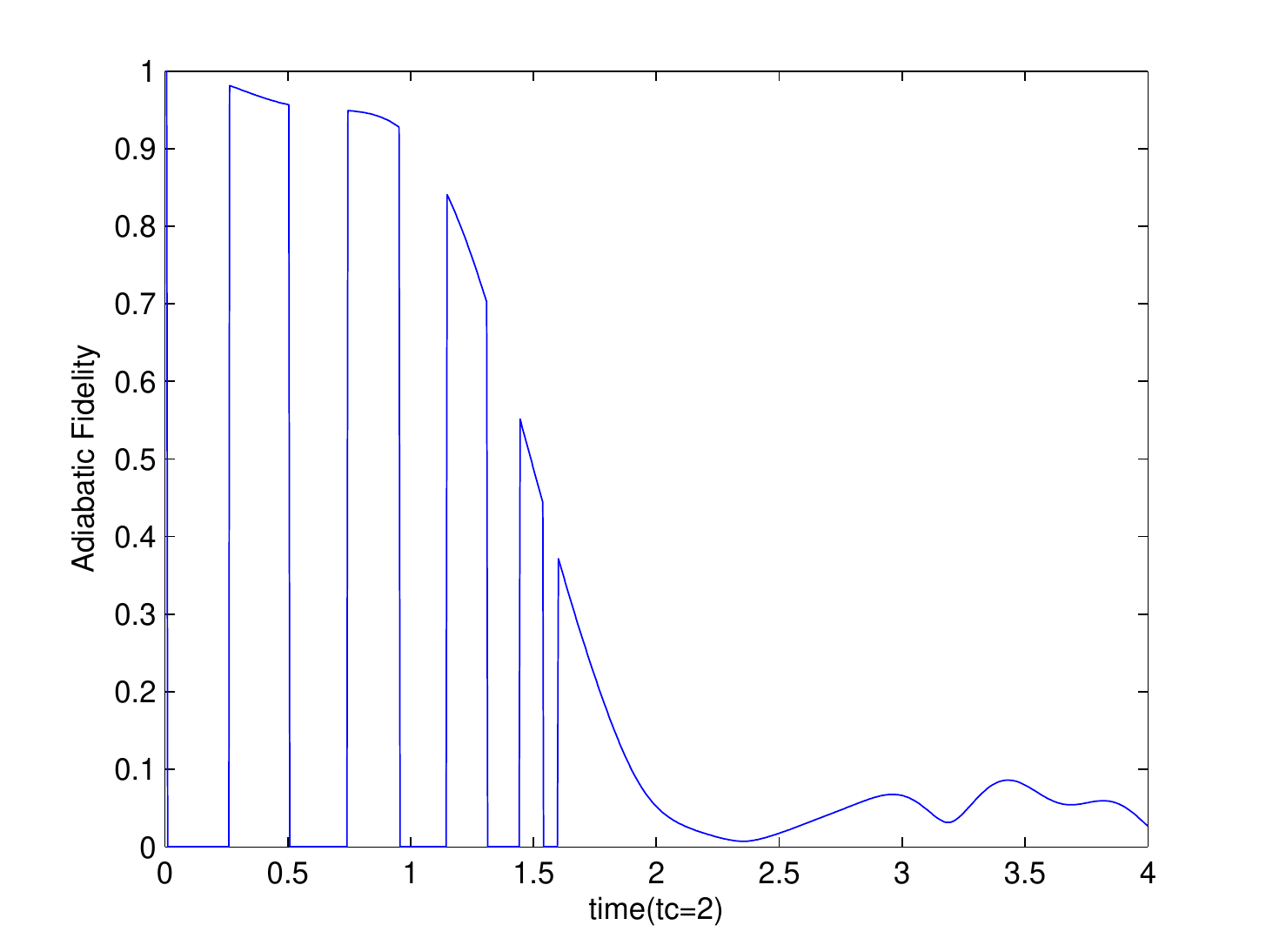}} \hspace{1cm}
\subfloat[][Parity for $N=17$ with $\mu_i \neq 0$.]{\includegraphics[width=0.4\linewidth]{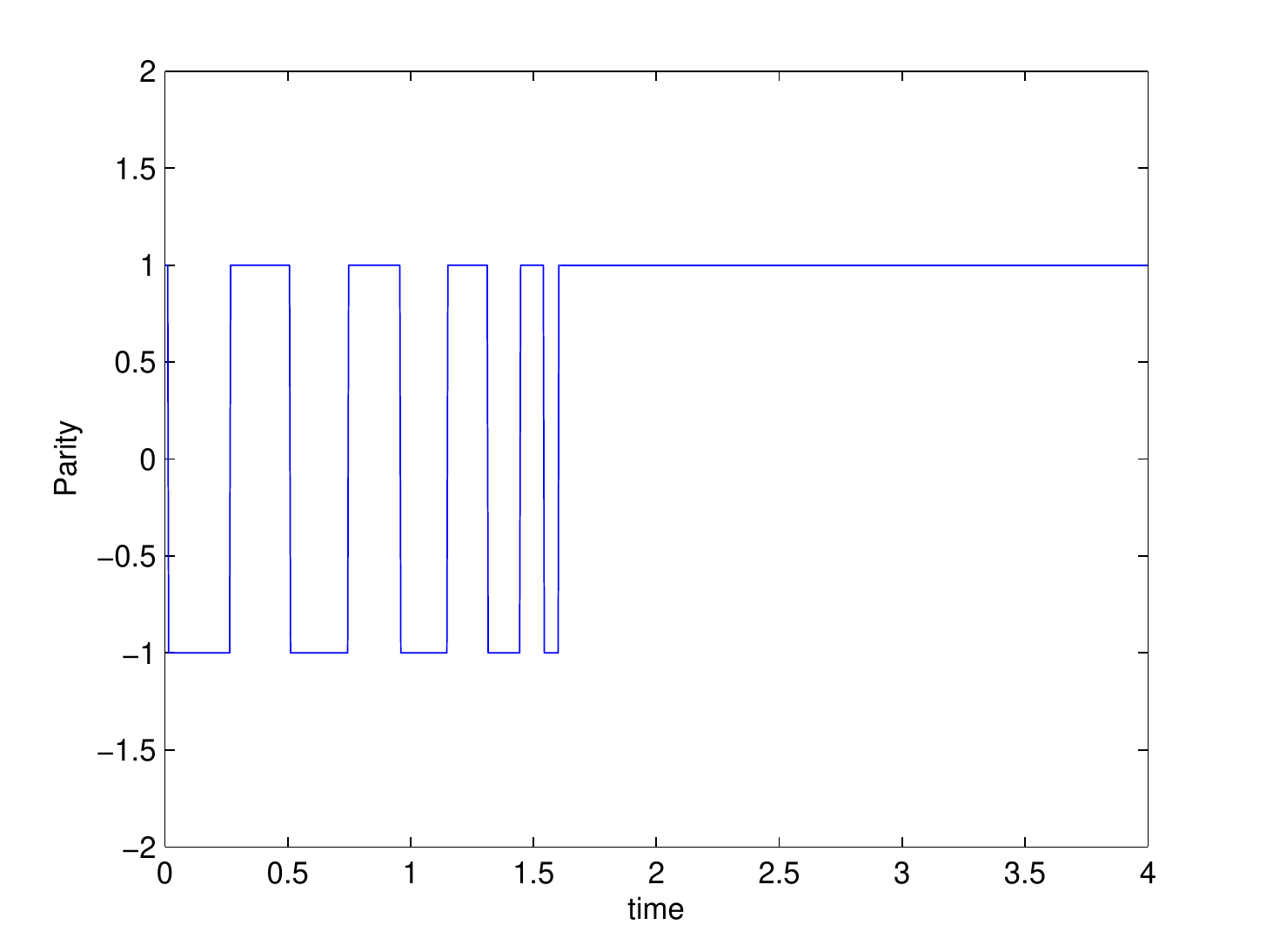}}
\caption{(Color online) Numerical results for (a) adiabatic fidelity $\mathcal{O}(t)$ and (b)parity of the
instantaneous ground state for an odd number of sites. In this case the system
has the same parity as the initial ground state on crossing the quantum
critical point (Figure (b)) and therefore has a non-vanishing overlap with it.}
\label{fig:OverlapOpen13}
\end{figure*}



Figures ~\ref{fig:OverlapOpen34} and ~\ref{fig:OverlapOpen13} show the numerical results for the
adiabatic fidelity $\mathcal{O}(t)$ along with the parity of the instantaneous ground state
for an open chain with an even and odd number of sites, respectively. The
case of the initial value $\mu_i=0$ for an odd number of sites will be
discussed later.

We can see from Figs. ~\ref{fig:OverlapOpen34} and ~\ref{fig:OverlapOpen13} that for both even
 and odd number of sites, the system starts in a particular fermion parity sector, and as it
moves within the topological phase the instantaneous ground state switches
parity regularly. On crossing the critical point it can either have opposite
parity from the initial state or the same parity, depending on the initial
parity sector. On the other hand, as we are dealing with parity
conserved systems, the state which is time evolved from the
initial ground state continues to have the same fermion parity. Thus the overlap
of the time evolved state with the
instantaneous ground state plunges to zero at times when the instantaneous parity
becomes opposite to the initial parity. We call these parity oscillations, which
occur for an open Majorana wire, as the parity blocking effect. The initial
ground state is blocked from having any non-zero overlap with the instantaneous ground
state for certain values of $\mu$. Finally, on crossing the quantum critical
point it becomes zero at all times if the parity is flipped from the initial
ground state; this is also a manifestation of parity blocking.
Hence, in Fig. ~\ref{fig:OverlapOpen34}, the case of an even number
of sites, the parities of the initial and final ground states are the opposite and
the system shows parity blocking for the entire topologically trivial phase, while in
Fig. ~\ref{fig:OverlapOpen13}, the parities are matched and there is some residual
overlap in the trivial phase.

\begin{figure}
\centering
\includegraphics[width=0.4\textwidth]{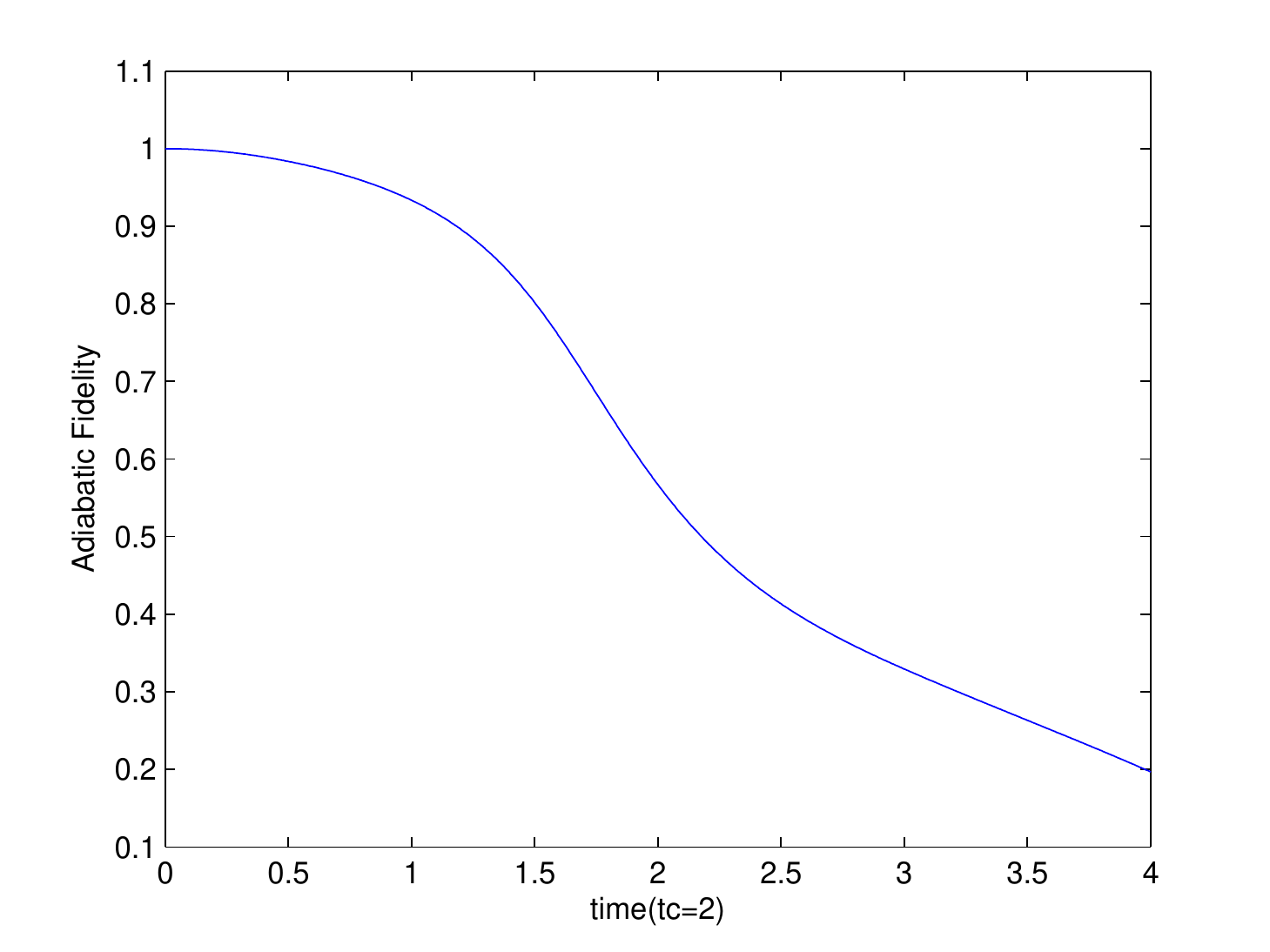}
\caption{(Color online) Numerical results for quenching with $\mu_i> 2\sqrt{1-\Delta^2}$. i.e outside the domain of oscillations as shown in the phase diagram. Here the nature of Majorana wave functions at the edges are purely decaying and their coupling would not have any oscillations, which would result in the ground state parity not switching as one sweeps through the parameter space. }
\label{fig:OverlapNoOsc}
\end{figure}

\textbf{Domain with no parity blocking:}
As we showed in Sec. ~\ref{sec:finitesize}, the oscillations in the parity do
not occur throughout the parameter space corresponding to the topological
phase. Namely, there are no oscillations if $\mu^2 > 4 (w^2 -\Delta^2)$.
This implies that there ought to be no parity blocking in the adiabatic
fidelity. We indeed see this in the numerical results shown in
Fig.~\ref{fig:OverlapNoOsc}.

\textbf{Fermion parity degeneracy for an odd number of sites:}
For an open chain with an odd number of sites, and for states which belong to
the odd fermion sector, the $\mu_i=0$ point is special in that it has two
degenerate ground states. To see this, let us define an operator
\beq C ~=~ i^{N(N-1)/2} ~a_1 a_4 a_5 a_8 \cdots , \eeq
where the last term on the right hand side is given by $a_{2N-1}$ if $N$
is odd and $a_{2N}$ if $N$ is even. We note that $C$ is both Hermitian
and unitary, so that $C^2 = I$. Recalling that $f_n = (1/2) (a_{2n-1} + i
a_{2n})$ and $f_n^\dagger = (1/2) (a_{2n-1} - i a_{2n})$, we find that $C$
generates the particle-hole transformation
\bea C f_n C &=& (-1)^{n+N-1} ~f_n^\dagger, \non \\
C f_n^\dagger C &=& (-1)^{n+N-1} ~f_n. \label{charge} \eea
This is a symmetry of the Hamiltonian in Eq. (\ref{ham1}) if $\mu = 0$.
We now note that the parity and charge-conjugation operators $P$ and $C$
satisfy $P C = (-1)^N C P$. Thus $P$ and $C$ anticommute if the number of
sites $N$ is odd. Since $P$ and $C$ both commute with $H$ if $\mu=0$,
every energy of a system with an odd number of sites will have a two-fold
degeneracy with the two eigenstates having opposite fermion parities.
(This can be shown as follows. If $|\psi_+>$ is an eigenstate of $H$ and $P$
with eigenvalues $E$ and $+1$ respectively, the relations $PC = -CP$ and
$HC = CH$ imply that $|\psi_-> = C|\psi_+>$ is an eigenstate of $H$ and
$P$ with eigenvalues $E$ and $-1$ respectively).
\begin{figure*}
\subfloat[][Partial blocking in adiabatic fidelity for different $N$.]{\includegraphics[width=0.3\textwidth]{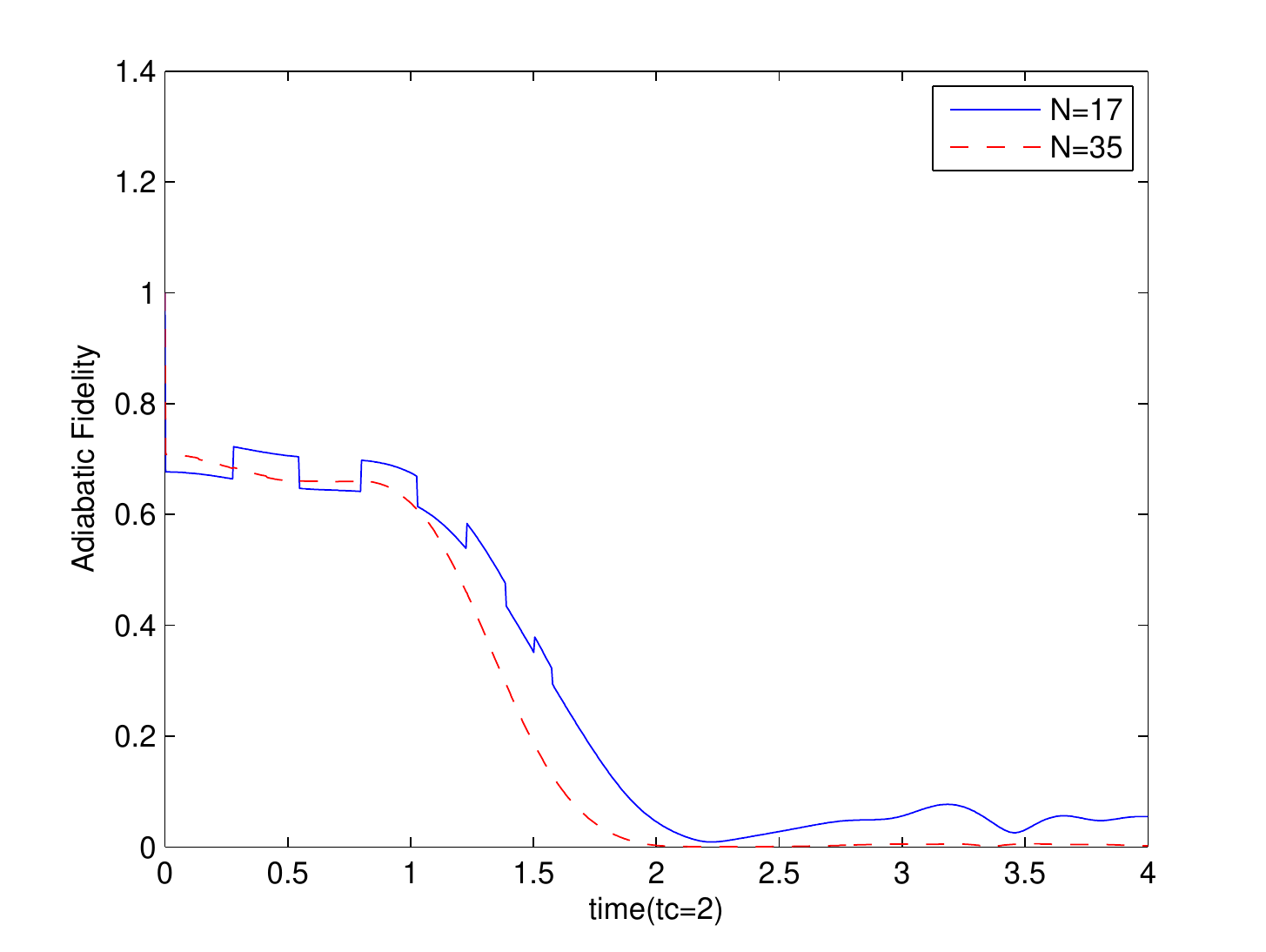}}\hfill
\subfloat[][No energy splitting at $t=0$.]{\includegraphics[width=0.3\textwidth]{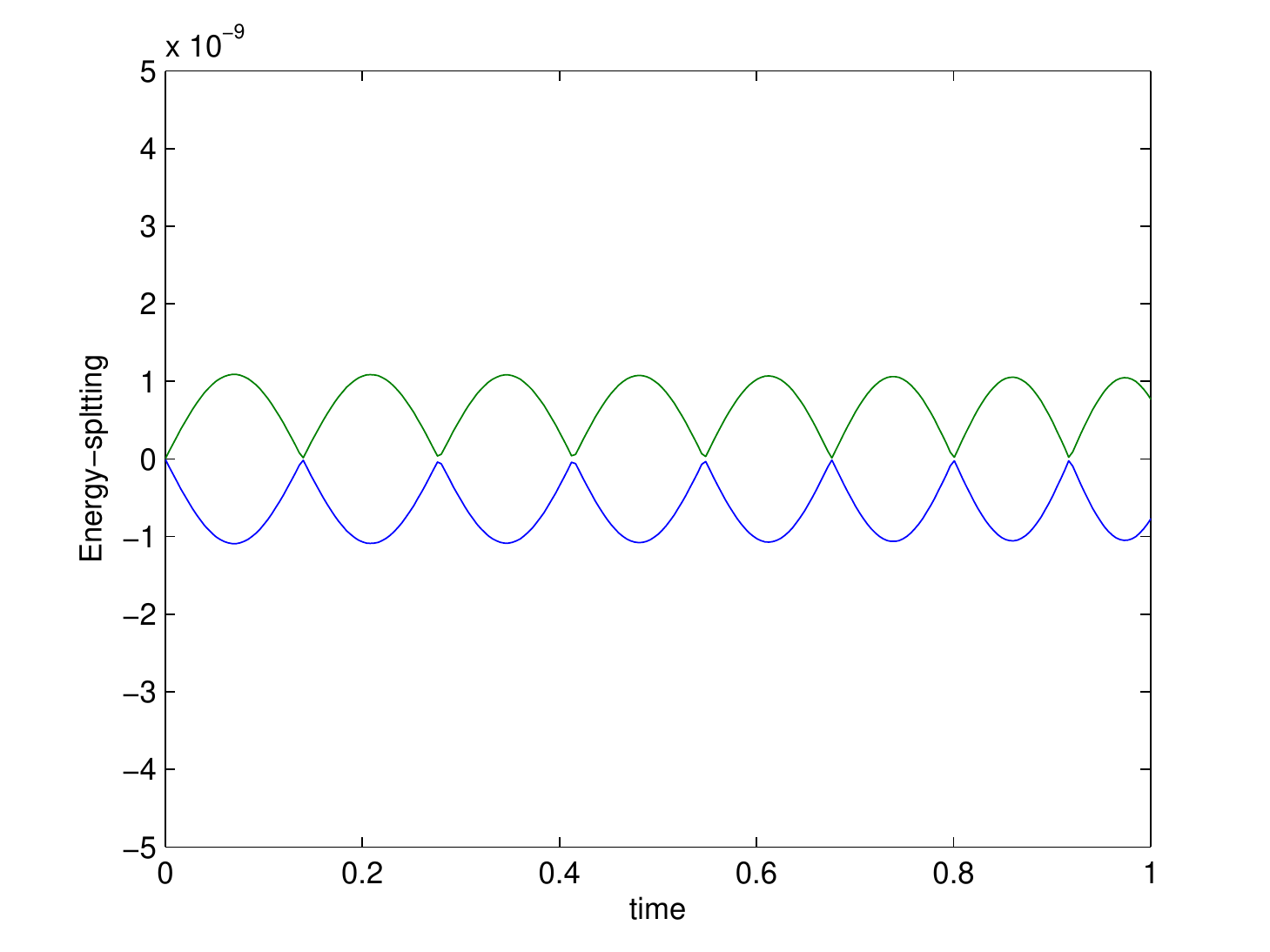}}\hfill
\subfloat[][Usual parity switching of the instantaneous ground state.]{\includegraphics[width=0.3\textwidth]{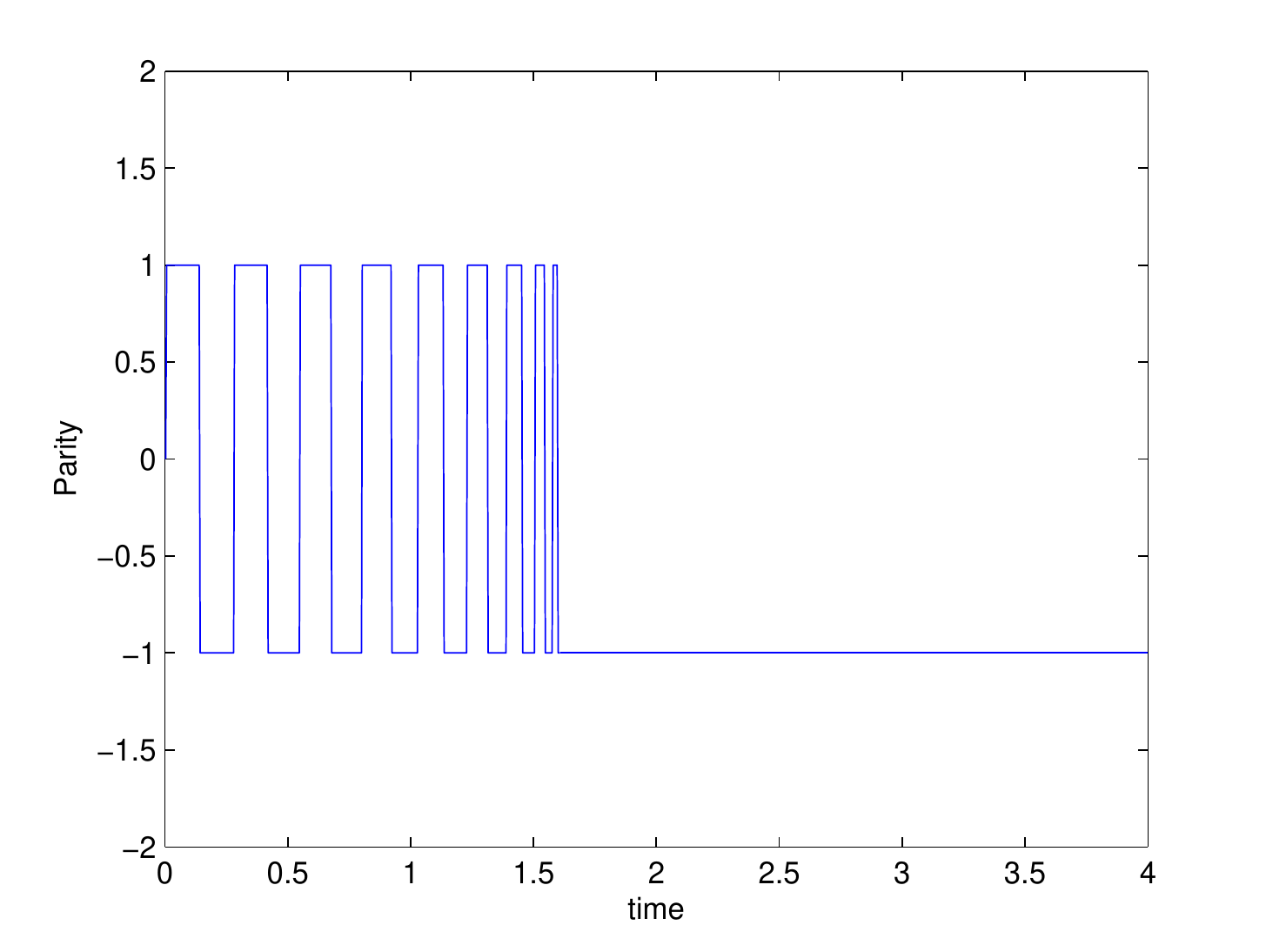}}
\caption{(Color online) Numerical results for quenching with $\mu_i=0$ for the odd
sector. This is the special case where the initial state is in a superposition of the odd and even parity states.Thus the time evolved states will not be completely 'parity blocked' but the amplitude of adiabatic fidelity will be reduced. As we go to smaller $N$ the splitting is exponentially enhanced and one can clearly see the effect of it in 'skewing' the superposition towards the state which
contributes to the ground state.}\label{fig:OverlapOpen35mu0}
\end{figure*}

Therefore, starting from $\mu_i=0$ means starting with parity states whose degeneracy is
not split. The time evolution of an arbitrarily chosen ground state would
therefore be that of a linear combination of both parity states. Even though the parity of the
instantaneous ground state would keep switching as one sweeps through the
topological phase, the overlap would be finite as the time evolved state
will be in a superposition of both parities. But the value of the adiabatic fidelity
will be smaller than in the parity blocked case as the amplitude is split
between the two superposed states. Figure ~\ref{fig:OverlapOpen35mu0} shows the results for this unique
case of quenching. We can clearly see in Fig. ~\ref{fig:OverlapOpen35mu0} that the splitting is zero
at time $t=0$. This change in the initial condition drastically affects the
evolution of the ground state and we do not see complete parity blocking in
this case.

\begin{figure*}
\subfloat[][]{\includegraphics[width=0.5\linewidth]{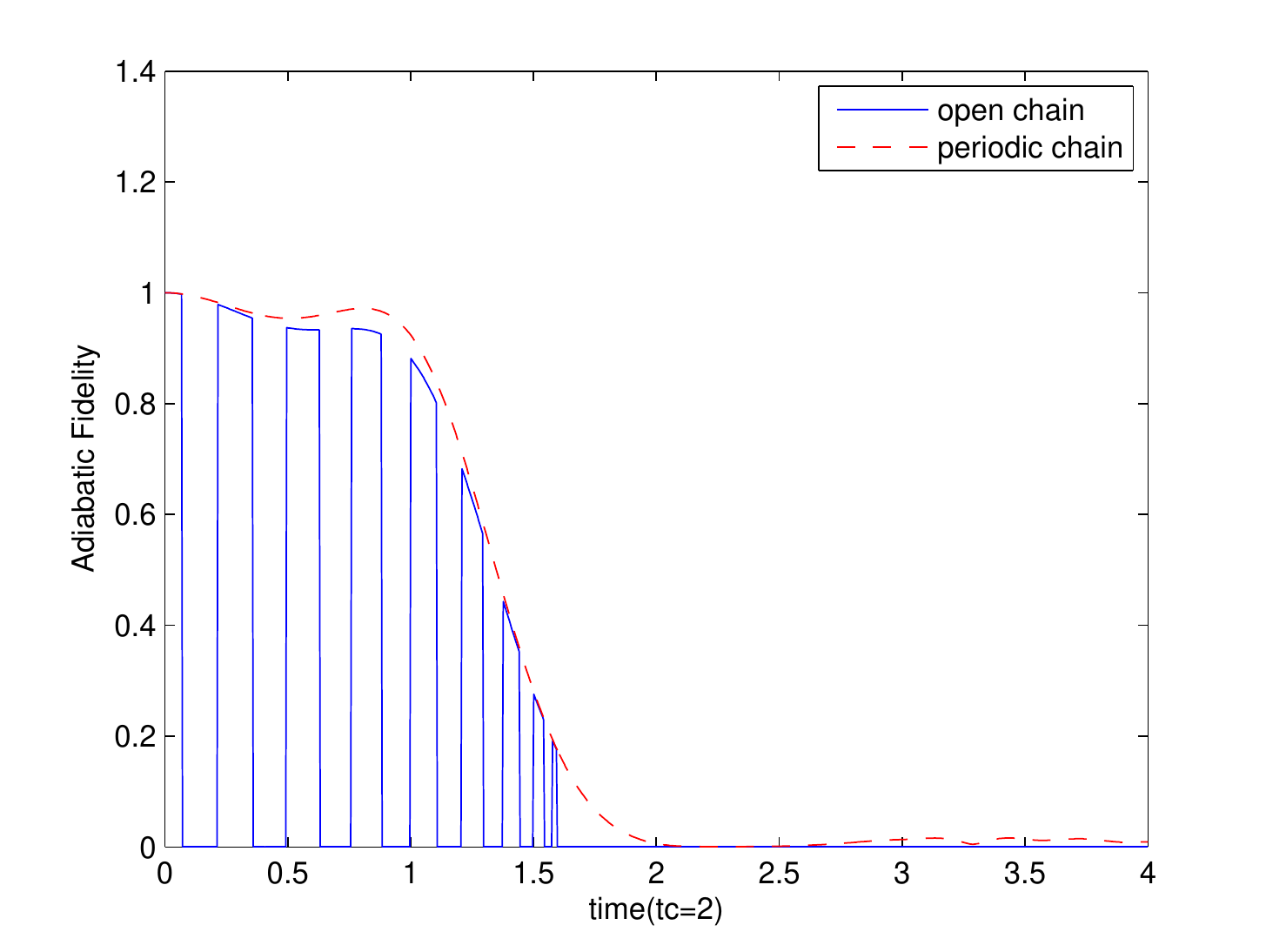}}\hfill
\subfloat[][]{\includegraphics[width=0.5\linewidth]{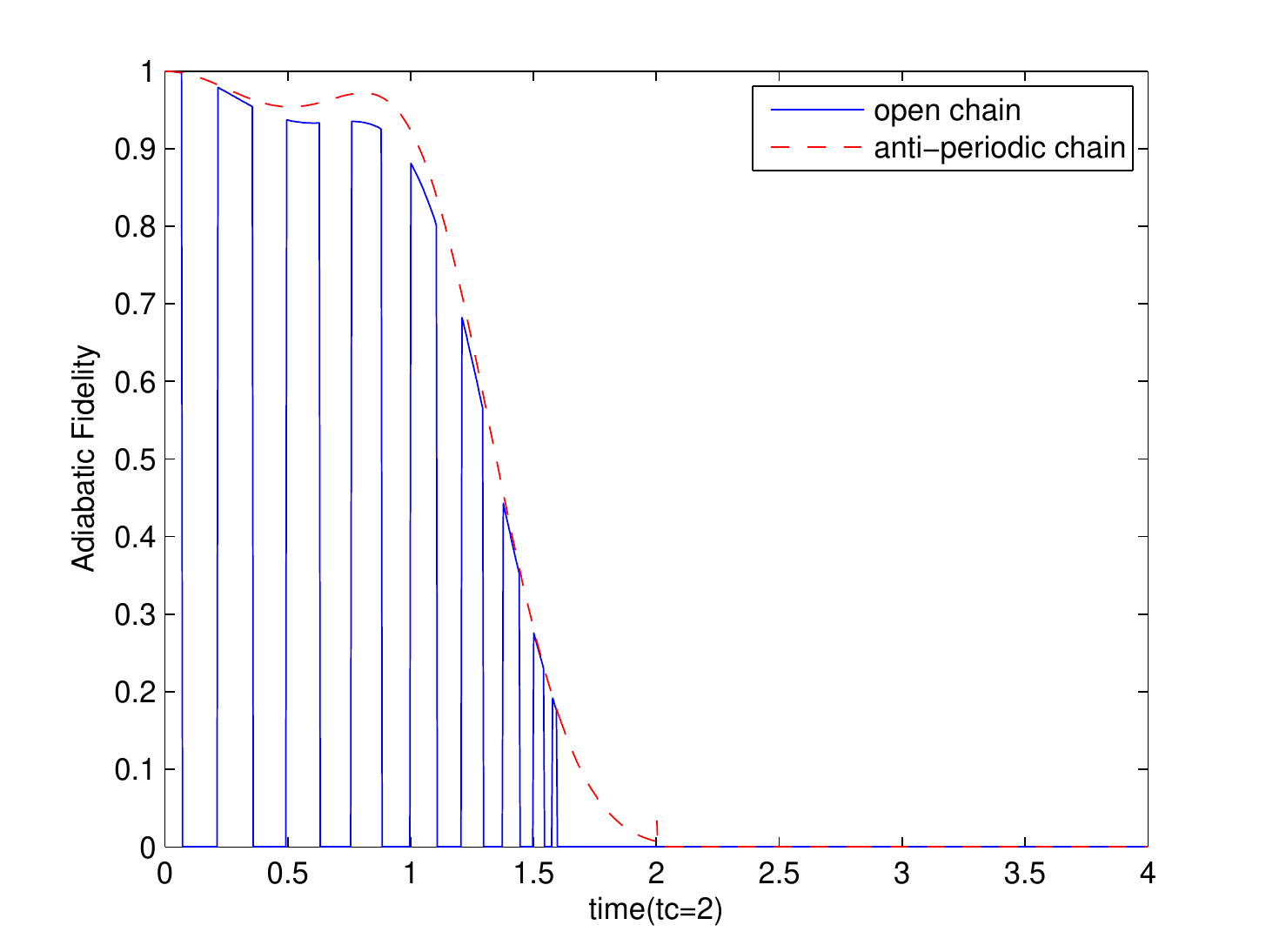}}
\caption{(Color online) Numerical calculation of $\mathcal{O}(t)$ for a closed chain with $34$ sites. The periodic and antiperiodic closed chains represent even and odd fermion sectors respectively. One can see that there is blocking in the second case, whereas a small amount of overlap persists in the first case after crossing the critical point. Also the envelope of the adiabatic fidelities for closed chains is compared with that of the open chain for the same number of sites. Even though there is no 'parity blocking' within the topological phase in the case of closed chain due to absence of the edge modes, the overall the behavior remains qualitatively the same.}
\label{fig:Overlapper}
\end{figure*}

\textbf{Comparison with a closed chain:}
The case of parity blocking in closed chains with periodic and antiperiodic
boundary conditions has been studied in detail in a previous
work~\cite{Kells14a}. While the previous analysis involved momentum modes,
simplifying the problem to a set of two-level Landau-Zener systems, the
real space formalism is easily extended here to compute all the quantities
for a closed chain. The numerical results are shown in Fig. ~\ref{fig:Overlapper}. Even though
there is parity blocking in the case of an open chain, we may still expect
the envelope of the adiabatic fidelity to be comparable with that of the
periodic case. In Fig.~\ref{fig:Overlapper}, we see a good match between the
envelopes in the two cases. 
For a closed chain, the final parity can flip from the initial state
depending on the boundary conditions. As can be seen, the parity does not
change for the periodic closed chain and therefore one
still has a finite overlap. The match between the envelopes in the open and
closed chain cases is very close. This suggests that overall the open chain
would also respect the Kibble-Zurek behavior for the defect production and
excitation density that was found in the closed chain case.
The crucial difference between open and closed chains is the parity blocking
and switching due to the coupling of the two Majorana end modes.

\subsection{Residual Energy}
\begin{figure*}[htp]
\centering
\subfloat[][Residual energy for an open chain with $N=36$ and $N=18$.]{\includegraphics[width=0.4\textwidth]{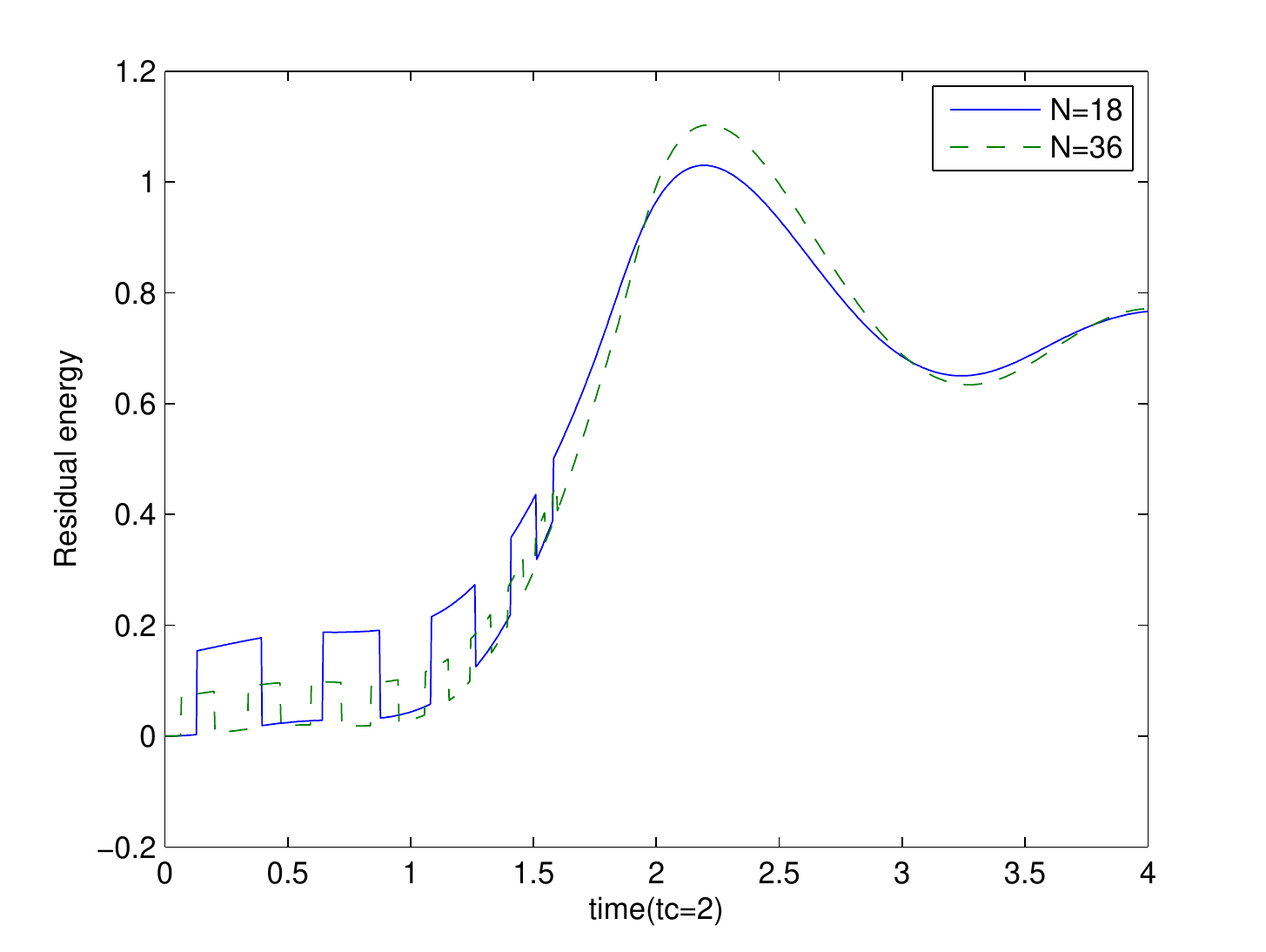}}\hspace{1cm}
\subfloat[][Residual energy for a periodic closed chain with $N=34$]{\includegraphics[width=0.4\textwidth]{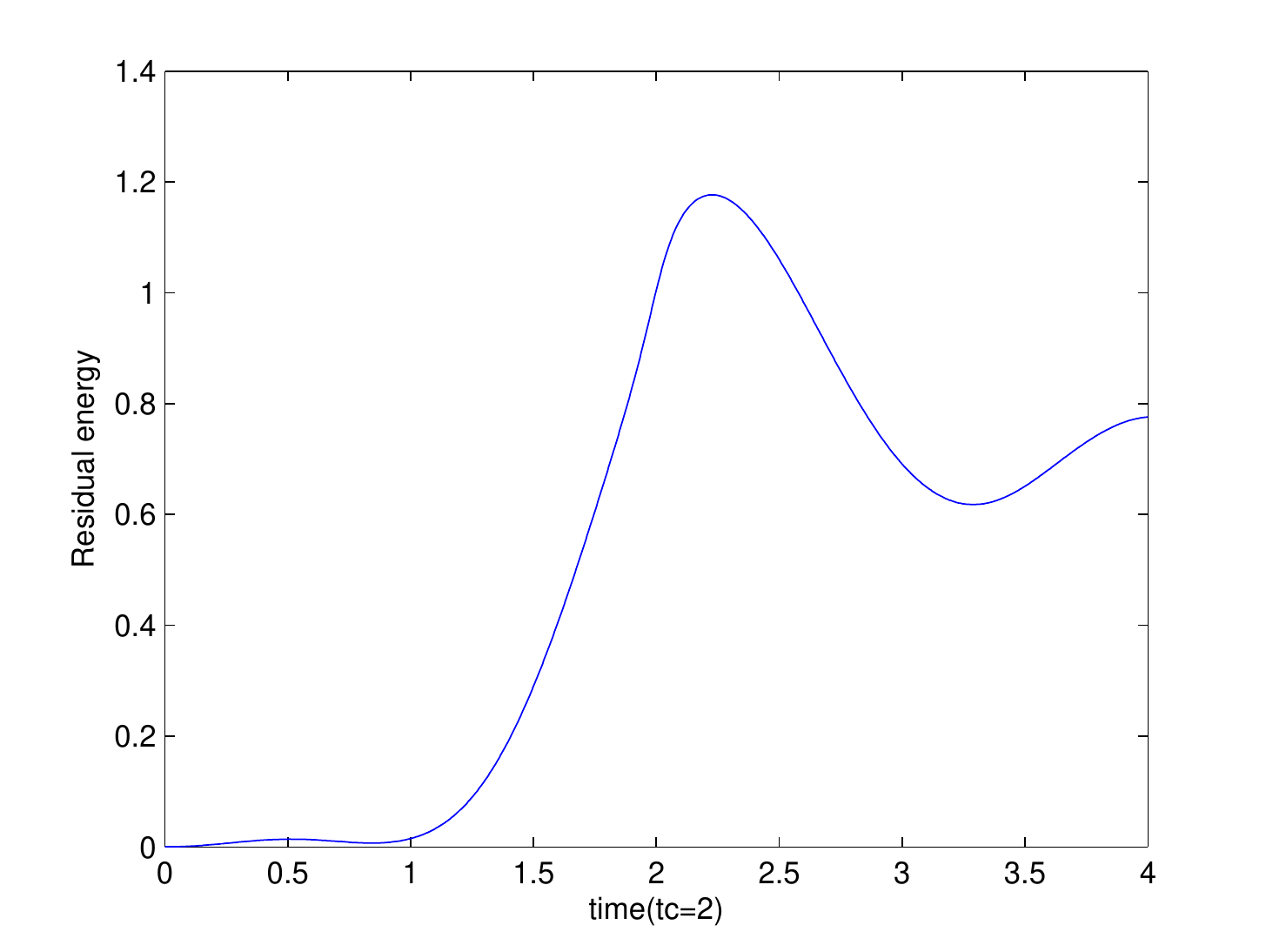}}
\caption{(Color online) Residual energy plots with the critical point occurring at $t=2$. One can notice in the case of open chain the oscillations before crossing the critical point, which arise due to the oscillation of mid-gap states.One can see that the steps arising due to the splitting scales inversely with the system size.}
\label{fig:EresOpen34}
\end{figure*}

The numerical results for the variation of the residual energy with time using
the full many-body formulation in Eq. (\ref{resen}) is shown in Fig.~\ref{fig:EresOpen34}
for both open and closed chains. The two cases have the same average behavior.
Both show a rapid increase in the energy of the system above the instantaneous many-body
ground state as they approach the critical point. This rapid rise is due to the system
falling out of equilibrium upon approaching the critical point and thus losing
adiabaticity. Far beyond the critical point, we find that the energy asymptotes to
a fixed average value. The scaling analysis of this quantity for the transverse field Ising chain has been studied numerically in ~\cite{Kolodrubetz12}.

\textbf{Effect of Majorana end modes:} The crucial difference between the open and
closed chains in Fig.~\ref{fig:EresOpen34} is the presence of the abrupt jumps at
small times in the case of the open chain. Reflecting the behavior of adiabatic fidelity, these
jumps correspond to switching back and forth between the ground state and excited
states due to parity blocking. Upon comparing the behavior of the residual energy
with the energy splitting and parity switching of Fig.~\ref{fig:SplittingOpen35},
we find that there is a complete match between the points at which the jumps
in the residual energy take place and the points where the parity switching
occurs. 

\textbf{Signatures of Loschmidt echoes:}
In addition to parity blocking, in our many-body system, we find evidence for the
Majorana-mode related physics found in previous work on single particle dynamics in quenching~\cite{Rajak14}. The Loschmidt echo studied in this work calculates the probability for an initial Majorana mode
to become a single-particle bulk excitation as a function of time as one sweeps through the critical point. If one varies $\mu(t)$ extremely slowly so as to be close to the adiabatic limit, the gap
between the mid-gap end states and bulk states scales as $1/N$ on approaching the critical point since the dynamical critical exponent is equal to 1.
The level spacing of the low lying bulk states also scales as $1/N$.
Hence the Loschmidt echo turns out to be a periodic function with period
$N$ as one quenches to or across the critical point.

\begin{figure*}[htp]
\centering
\subfloat[][$N=18$]{\includegraphics[width=0.4\linewidth]{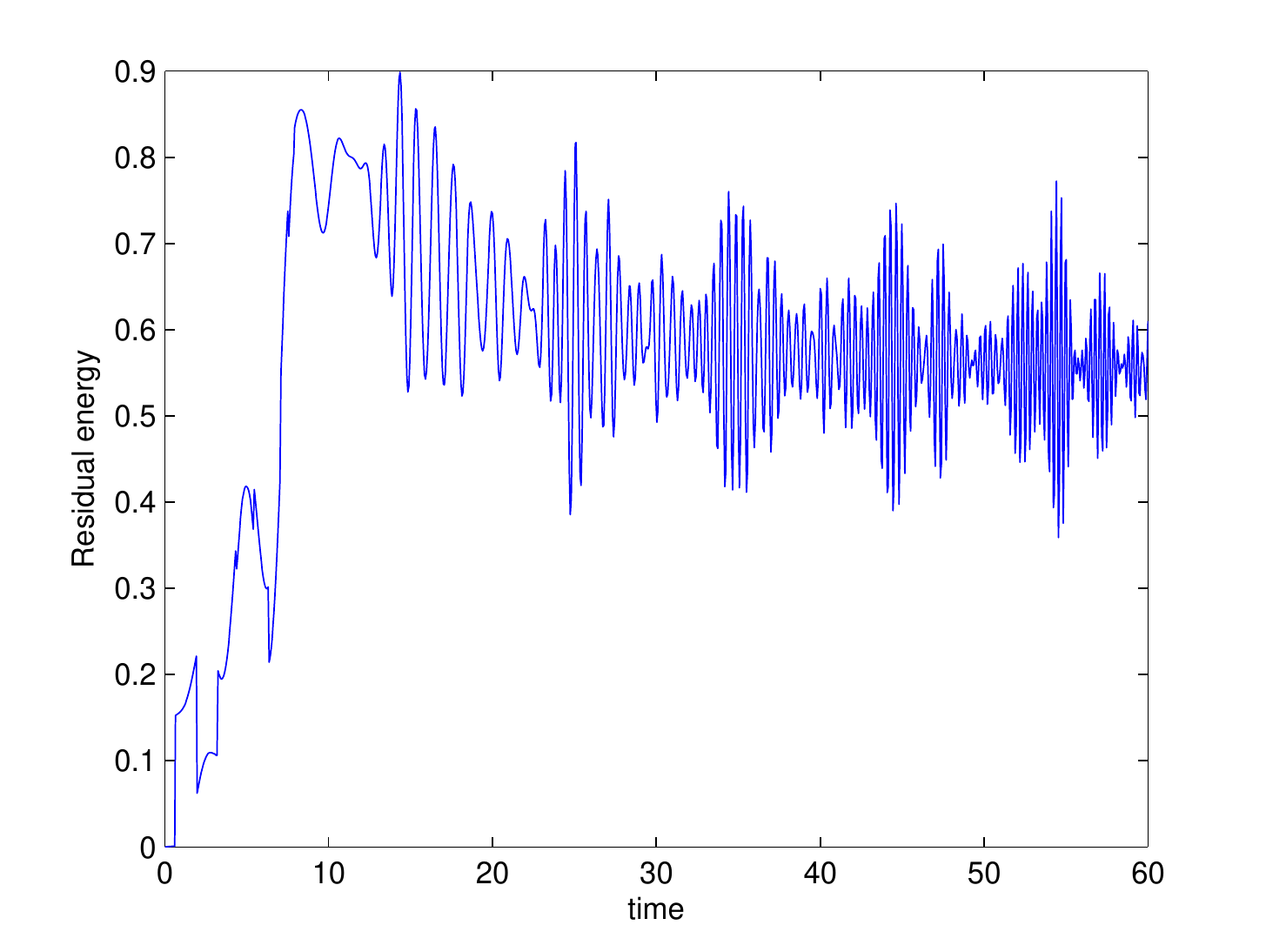}} \hspace{1cm}
\subfloat[][$N=36$]{\includegraphics[width=0.4\linewidth]{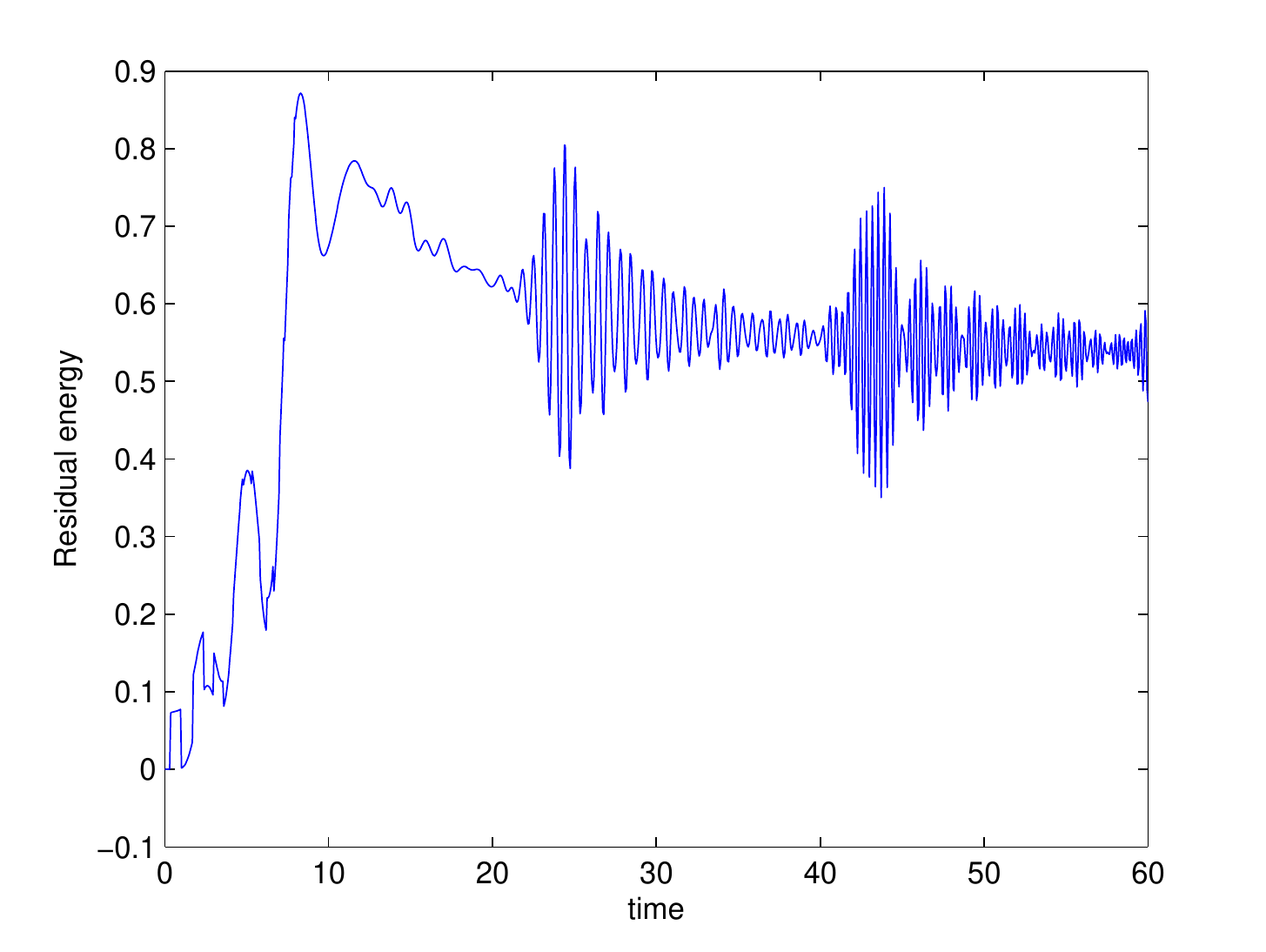}}
\caption{(Color online) The residual energy plots for a very small quench rate which is
nearly adiabatic. In this case one can see the periodic recurrence of
excitations at times after crossing the critical point. The period is doubled if we double the system size.}
\label{fig:Eres18v1}
\end{figure*}

In our full-fledged many-body treatment, these echoes appears as `chirps' of excitations whose occurrence has a period of $N$. These excitations contribute to the overall energy of the state. Note that this is true only if the quench rate is extremely slow. As shown in Fig.~\ref{fig:Eres18v1}, one can see these oscillations in the numerical results at a low quench rate like $1/\tau=0.1$.
The frequency of occurrence of these chirps in the excitations is indeed
halved when the system size is doubled. These chirps also appear in the
adiabatic fidelity.

To summarize, open and closed chains broadly show similar average behaviors
as expected for quench dynamics, However, both in the adiabatic fidelity and
residual energy, distinct non-analytic features arise in the form of jumps
only for an open chain, and these can be attributed directly to the presence
of end modes and their associated fermion parity.

\section{Discussion}
\label{sec:discussion}

In conclusion, our study of non-equilibrium behavior in finite-sized Majorana
wires demonstrates that the presence of topological order can dramatically
alter quench dynamics. Previous work involving Majorana wires having periodic
boundary conditions brought to light the notion of topological blocking in
tuning between different quantum phases ~\cite{Kells14a}. In contrast,
we have seen here that the coupling between Majorana end modes and the
associated ground state parity flips as a function of a tuning parameter
gives rise to a more drastic manifestation of topological blocking due to a
succession of switches between topological sectors within a single topological
phase. As a result, some common measures studied in the quench dynamics
literature, such as wave function overlaps between time-evolved states and
instantaneous ground states (the adiabatic fidelity), and residual energies,
show a series of non-analytic structures in the form of characteristic
jumps which are not observed in standard Kibble-Zurek physics.

Our work has shown that there is a much richer texture in the phase diagram
of the Kitaev chain or the Majorana wire than has been presented earlier.
The circle which separates the regions of oscillating and purely decaying
wave functions of the Majorana end modes exists in the thermodynamic limit.
In addition, we have shown that a coupling between the end modes, due to
a finite length of the chain, leads to further divisions within the circle
in the form of ellipses, each division corresponding to a particular fermion
parity of the ground state. Although the energy splitting between the
Majorana modes goes to zero as we increase the chain length, the number of
fermion parity switches increases linearly with the length. This has a
dramatic consequence for the adiabatic fidelity under a quench, namely,
parity blocking occurs more frequently as we increase the system size.
(This is very different from conventional finite size effects which
typically vanish in the thermodynamic limit). We therefore see that
parity blocking is not merely a finite size effect, but is a relevant
manifestation of the physics of Majorana modes and their topological
nature in any real system.

This study shows that quench dynamics serves well as a probe of topological
order. While blocking features need not be unique to topological systems in
that quantum invariants in other systems can possibly have similar effects,
they are necessary conditions under appropriate circumstances (for instance,
open boundary condition in the case studied here). Moreover, unlike in most
other systems, such as ferromagnets having local order, we expect this blocking
phenomenon to be robust against local perturbations. In the case of a
finite-sized Majorana wire, the succession of parity switches associated with
topological sectors is a crucial aspect of topological order; while studies of
the static behavior have been extensive (see, for instance,
~\cite{Burnell14,Zazunov13,Bonderson11,Hassler11}), here we have performed
the nearly unexplored study of their effect on quench dynamics.

In fact, the issue of parity forms the basis of several discussions and proposals
for Majorana wires, particularly in light of the potential experimental discovery
of isolated Majorana end modes and their implications for topological braiding
and quantum computing. Several schemes involve changing the on-site chemical
potential at specific locations as a means of manipulating and dynamically moving
the isolated end modes. A popular study regarding the end modes is the fractional
Josephson effect
(see, for instance, ~\cite{Kitaev01,Fu09,Law11,Kwon04,Beenakker13, Crepin14}),
which involves parity switches between two finite-sized Majorana wires
connected to each other at their ends and their effect on Josephson physics
(in principle, other zero energy end bound states could mediate such an effect).
Our study here is highly relevant to these lines of investigation and it provides
a dynamic quantum many-body formulation that goes beyond quasi-static approximations.

While our study primarily aims to understand the effects of parity switching on
issues typically studied in the literature on quench dynamics, an experimental
setup probing the predictions would be remarkable. While the arena of cold atomic
gases is more ideally suited for measurements of residual energy and adiabatic fidelity, realizing topological order in these systems is still in its initial phases \cite{Mazza12}.
In the setting involving spin-orbit coupled wires, where the isolated end modes have
potentially been observed, in principle, ground state parity switches can be
observed by coupling the wire to another system. For instance, a possible read-out
could involve tunnel-coupling to a quantum dot or STM tip (see, for example,
~\cite{Flensberg12,Lee14}).
Further studies would involve pinpointing ways of measuring the behavior of adiabatic fidelity predicted here
in such a setup.

Finally, the study of topological/parity blocking in quench dynamics presented here
and the associated quantum many-body formulation offer wide scope for further
exploration. Several aspects of this initial study require more detailed investigation,
for instance, more involved studies of system size, and further connections with
Kibble-Zurek scaling and single-particle physics, including anomalous scaling due to
boundary effects and appearance of Loschmidt echoes. Oscillations have been found in the derivative of the Renyi entropy with respect to the chemical potential in \cite{Ming15} and it may be interesting to see if there are such effects related to the oscillations in the ground state parity. A host of open issues related to topological blocking in Majorana wires include constraints on thermalization imposed by topological order, effects of external potentials, such as quasiperiodic
potentials and disorder, and higher dimensional analogs, such as the Kitaev
honeycomb model.

\verb"Acknowledgements"
We thank Bryan Clarke, Graham Kells and Jeffrey Teo for illuminating 
conversations. This work is supported by the National Science Foundation under the grants DMR 0644022-CAR (SH) and DMR-0906521 (VS) , the U.S. Department of Energy, Division of Materials Sciences under Award No. DE-FG02-07ER46453 (SV), and by the Department of Science and Technology, India: SR/S2/JCB-44/2010. For 
their hospitality during the course of this work, D. S. gratefully 
acknowledges the Department of Physics, University of Illinois at 
Urbana-Champaign and S.V. the Aspen Center for Physics.


\section*{Bibliography}
\bibliographystyle{iopart-num}
\bibliography{NJParxiv}{}

\appendix
\section*{Appendix}
\setcounter{section}{1}
\subsection{Calculation of adiabatic fidelity $\mathcal{O}(t)$}
Let us start with a general Hamiltonian which is quadratic in Majorana
operators $a_j$,
\beq H=i\sum\limits_{i,j=1}^{2N} a_i M_{ij} (t) a_j. \eeq

The antisymmetric Hermitian matrix $iM$ has real eigenvalues which
come in pairs $\pm \lambda_j$ with corresponding eigenvectors $x_j$ and
$x^*_j$, which are orthonormal to each other. We define a set of linear
combinations of Majorana operators in terms of these eigenvectors as
\bea b_j^{\dagger}(t) &=& \frac{1}{\sqrt{2}}\sum_{i=1}^{2N} ~(x^T_j)_i
a_i, \non \\
b_j(t) &=& \frac{1}{\sqrt{2}} \sum_{i=1}^{2N} ~(x^{\dagger}_j)_i a_i. \eea
In terms of a $(2N)$-component vector $\bar{b}=(b_1,b_2, \cdots, b_N,
b^{\dagger}_1, \cdots, b^{\dagger}_N)^T$, Eq. (A2) can be expressed as a
linear transformation
\beq \bar{b}(t)=B(t)\bar{a}. \eeq
The rows of the $(2N)$-dimensional matrix $B$ are the eigenvectors $x^T_j$ and
$x^{\dagger}_j$, and $B$ belongs to the unitary group $U(2N)$ with $det(B)=
\pm 1$. In terms of $b_j$, the Hamiltonian (A1) becomes, up to a constant,
\beq H=4\sum\limits_{j=1}^N \lambda_j b^{\dagger}_j b_j. \eeq
If $|\Psi(0)\rangle$ is the initial ground state of $H$ at $t=0$, then $b_j(0)
|\Psi(0)\rangle=0$. The instantaneous ground state $\ket{\psi_{ins}(t)}$ of
$H(t)$ is annihilated by $b_j(t)$, namely, $b_j(t) \ket{\psi_{ins}(t)}=0$.

We now want to find the adiabatic fidelity $\langle \Psi(t)| \psi_{ins}(t)\rangle$. Let
us examine the operators which annihilate the time evolved state
$\ket{\Psi(t)}$ and some of their properties. Let us say $\beta_j(t)
\ket{\Psi(t)}=0$. We want to put the information of the time evolution with
$H(t)$ into $\beta_j(t)$. The time evolution of the Majorana operators $a_j$
is given by their Heisenberg equations of motion:
\beq \frac{da_j(t)}{dt}=-i[H(t),a_j(t)]= -4 \sum\limits_{k=1}^{2N}
M_{jk}(t)a_k(t). \eeq
In terms of a $(2N)$-component vector $\bar{a}$, the solution of the equation
$d\bar{a}(t)/dt =-4M(t)\bar{a}(t)$ is given by the evolution operator
\beq \bar{a} (t)= S(t,0) \bar{a} (0). \eeq
Here $S(t,0)=\mathcal{T}exp(-4\int_0^t M(t')dt')$ which can be calculated
numerically for a given $M(t)$.
It can now be shown that $\bar{\beta}(t)=B(0)\bar{a}(t)= B(0)S(t,0)
\bar{a}(0)$, where $\bar{\beta}(t)$ is the $(2N)$-component vector comprising
of $\beta_j(t), \beta^{\dagger}_j(t)$ similar to $\bar{b}(t)$. Therefore the
relation between $\bar{\beta}(t)$ and $\bar{b(t)}$ is given by
\beq \bar{\beta}(t)=B(0)S(t,0)[B(t)]^{-1}\bar{b} (t). \eeq
The key idea underlying the calculation in real space is to express all the
quantities of interest in terms of those which can be calculated
numerically. We can see that given the form of the initial Hamiltonian $H(0)$
and the time-dependent $H(t)$, the quantities $B(0)$, $B(t)$ and $S(t,0)$ can
be easily computed. Given these and the annihilation operators of the ground
states, we will now derive the final expression for the adiabatic fidelity.

Consider the Fock space of $2^N$ states $\ket{\phi_a}$, $a=1,2, \cdots, 2^N$.
Their fermionic occupation numbers are given by $\beta ^{\dagger}_j(t)
\beta_j(t)=0$ or $1$. For a $\ket{\Psi(t)}$ belonging to this Fock space,
$\beta_j(t) \ket{\Psi(t)}=0$; hence
$\beta^{\dagger}_j(t)\beta_j(t)\ket{\Psi(t)}= 0$
for all $j$. We now define the following operators $L_j$
\bea L_j&=& \beta_j(t) ~~{\rm for}~~ 1\leq j \leq N, \\
&=& \beta^{\dagger}_{2N+1-j} ~~{\rm for}~~ N+1 \leq j \leq 2N. \eea
In the Fock space, $\ket{\Psi(t)}$ is the only state which is not
annihilated by the product $L_1 L_2 \cdots L_{2N}$. In fact,
\beq L_1 L_2 \cdots L_{2N} \ket{\Psi(t)}=\ket{\Psi(t)}. \eeq
We will make use of this fact to reduce the calculation in the
$2^N$ dimensional Fock space to a matrix computation in $2N$ dimensions as
follows. Consider the quantity
\bea && \non \bra{\psi_{ins}(t)}L_1 L_2 \cdots L_{2N} \ket{\psi_{ins}(t)}
\non \\
&& = \bra{\psi_{ins}(t)} L_1 L_2 \cdots L_{2N} \sum\limits_{a=1}^{2^N}
\ket{\phi_a}\langle \phi_a \ket{\psi_{ins}(t)} \non \\
&& =\bra{\psi_{ins}(t)} L_1 L_2 \cdots L_{2N} \ket{\Psi(t)}\langle \Psi(t)|
\psi_{ins}(t)\rangle \non \\
&& =|\langle \psi_{ins}(t)\ket{\Psi(t)}|^2. \eea
This can be simplified further using a form of Wick's theorem given in
Ref.~\cite{Bravyi12}, in terms of a $(2N)$-dimensional antisymmetric
matrix $A$ defined as
\begin{eqnarray*}
A_{jk} &=& \bra{\psi_{ins}(t)}L_j L_k\ket{\psi_{ins}(t)} ~~{\rm for}~~ j<k,
\non \\
&=& -\bra{\psi_{ins}(t)}L_k L_j\ket{\psi_{ins}(t)} ~~{\rm for}~~ j>k, \non \\
&=& 0 ~~{\rm for}~~ j=k. \end{eqnarray*}
This matrix can be calculated using the relation between $\bar{\beta}(t)$ and
$\bar{b}(t)$ and the fact that $b_j(t)\ket{\psi_{ins}}(t)=0$. Finally we get
\bea \non
\bra{\psi_{ins}(t)} L_1 L_2 \cdots L_{2N}\ket{\psi_{ins}(t)}=Pf(A)=|\langle
\psi_{ins}(t)\ket{\Psi(t)}|^2. \eea
Since the Pfaffian is given by $Pf(A)=\pm\sqrt{det(A)}$, we see that the adiabatic fidelity is
\beq |\langle \psi_{ins}(t)\ket{\Psi(t)}|= |det(A)|^{1/4}. \eeq

Now we need to calculate the matrix elements $A_{jk}$. Since $A$ is
antisymmetric, we need to calculate only the elements for $j<k$. These are
given by
\begin{eqnarray*}
&& A_{jk} \non \\
&=& \bra{\psi_{ins}(t)}\beta_j \beta_k\ket{\psi_{ins}(t)} ~~{\rm for}~~
j\leq N,k\leq N, \\
&=& \bra{\psi_{ins}(t)}\beta_j \beta^{\dagger}_{2N+1-k}\ket{\psi_{ins}(t)}
~~{\rm for}~~ j\leq N,k>N, \\
&=& \bra{\psi_{ins}(t)}\beta^{\dagger}_{2N+1-j} \beta^{\dagger}_{2N+1-k}
\ket{\psi_{ins}(t)} ~~{\rm for}~~ j\leq N,k>N. \end{eqnarray*}
We need to evaluate each of these terms. We introduce $G(t)= B(0)S(t,0)
B^{-1}(t)$ so that $\beta(t)=G(t)b(t)$. Consider the first case
$j\leq N,k\leq N$:
\bea
&& \bra{\psi_{ins}(t)}\beta_j \beta_k\ket{\psi_{ins}(t)} \non \\
&=& \sum\limits_{m,n} \bra{\psi_{ins}(t)}G_{jm}\bar{b}_m G_{kn}\bar{b}_n
\ket{\psi_{ins}(t)} \non \\
&=& \sum\limits_{m,n} G_{jm}D_{mn}G_{kn}. \eea
The matrix D is given by
\begin{eqnarray*}
&& D_{jk} ~=~ \bra{\psi_{ins}(t)}\bar{b}_j \bar{b}_k\ket{\psi_{ins}(t)} \non \\
&=& \bra{\psi_{ins}(t)}b_j b_k \ket{\psi_{ins}(t)}=0 ~~{\rm for}~~ j\leq N,k
\leq N, \non \\
&=& \bra{\psi_{ins}(t)}b_j b^{\dagger}_{k-N}\ket{\psi_{ins}(t)}=\delta_{j,k-N}
~~{\rm for}~~ j\leq N,k > N, \non \\
&=& \bra{\psi_{ins}(t)}b^{\dagger}_{j-N} b_k\ket{\psi_{ins}(t)}=0 ~~{\rm for}~~
j> N,k \leq N, \non \\
&=& \bra{\psi_{ins}(t)}b^{\dagger}_{j-N} b^{\dagger}_{k-N}\ket{\psi_{ins}(t)}
=0 ~~{\rm for}~~ j> N,k > N. \end{eqnarray*}
Using this fact, we get
\bea 
&& \bra{\psi_{ins}(t)}\beta_j \beta_k\ket{\psi_{ins}(t)} \non \\
&=& \sum\limits_{m\leq N} G_{jm}(t)G_{k,m+N}(t) \non \\
&=& (\mbox{first half of j-th row of G})\times (\mbox{second half of k-th row of G})^T. 
\eea
Similarly the other elements of $A_{jk}$ can be found and numerically
evaluated as a function of time. Once we have the matrix $A$, the adiabatic fidelity
is simply related to its determinant.

\subsection{Parity in a two-site problem}
For an open chain, we saw in Sec.~\ref{sec:finitesize} that the overall parity of the ground
state is decided by the fermion parity of the split energy states arising
from the overlap of the Majorana end modes. We will consider here the effective
Hamiltonian for such a system and illustrate how $det(B)$ determines the parity
of the system. The effective Hamiltonian for the coupled Majoranas is given by
\beq H_f=i 2J a_1 a_{2N}. \eeq
The eigenvalues are given by $\lambda = \pm J$ and the eigenvectors are
$x$ and its conjugate $x^*$, where
\beq x=\frac{1}{\sqrt{2}}{1 \choose -i}. \eeq
Using these eigenvectors we can construct the matrix $B$ which transforms
the Hamiltonian into the canonical form.
\[ B=\frac{1}{2} \left[ {\begin{array}{cc}
1 & i \\
1 & -i \\
\end{array} } \right] \]
Now suppose that $J$ changes sign; then the eigenvalues are flipped and the
rows in $B$ are also flipped. The sign of the determinant of a matrix
changes when two of its rows are interchanged. From
Eq. (\ref{bBa}) we see that as the value of $J$ changes sign, the energy
corresponding to the state with a particular fermion parity changes. The
sign of $det(B)$ precisely tracks this flip in the parity of the energy
level which contributes to the ground state.

One can also see this from the calculation of the Pfaffian of the Hamiltonian for the two-site problem.
The Hamiltonian in the Majorana basis is given by:
\[ \left( \begin{array}{cccc}
0                &  -i \mu/2       & 0              & i(-w+\Delta)/2 \\
i\mu/2           &  0              & i(w+\Delta)/2  & 0  \\
0                & -i(w+\Delta)/2  & 0              & -i\mu/2  \\
-i(-w+\Delta)/2  & 0               & i\mu/2         & 0\end{array} \right)\]

The Pfaffian of this Hamiltonian is given by
\begin{equation}
Pf(H)=\frac{\mu^2}{4}-\frac{w^2-\Delta^2}{4}
\end{equation}
From this expression the Pfaffian changes its sign at $\frac{\mu^2}{4}=\frac{w^2-\Delta^2}{4}$. One can see that this is precisely the condition we have in Eq. (\ref{ellipse}) for $N=2$. As shown in Kitaev's paper ~\cite{Kitaev01} we have
the condition:
\begin{equation} P(H)= sgn[Pf(H)]=sgn[det(B)]. \end{equation}

\subsection{Calculation of residual energy}
The residual energy is defined as
\beq E_{res} = [\langle \Psi(t)| H(t)| \Psi(t)\rangle - E_G(t)]/ |E_G(t)|, \eeq
where $E_G(t)$ is the instantaneous ground state energy. Let us calculate the
first term in the expression, $\langle H(t)\rangle$. The matrix $B(t)$
transforms the time-dependent Hamiltonian to the canonical form
\bea H(t)&=& 4\sum\limits_{j=1}^N \lambda_j(t) b^{\dagger}_j(t) b_j(t)-2\sum
\limits_{j=1}^N \lambda_j(t) \non \\
&=& 4\sum\limits_{j=1}^N \lambda_j(t) \bar{b}_{N+j}(t) \bar{b}_j(t)-2\sum
\limits_{j=1}^N \lambda_j(t). \eea
Therefore, $\langle H(t)\rangle =4 \sum\limits_{j=1}^N \lambda_j \bra{\Psi(t)}
\bar{b}_{N+j}(t) \bar{b}_j(t)\ket{\Psi(t)} -2\sum\limits_{j=1}^N \lambda_j(t)$.
{}From Eq. (A7) and using the definition $G(t)= B(0)S(t,0)B(t)^{-1}$ from the
last section, we obtain
\beq \bar{b}_i(t)=\sum\limits_j G_{ij}^{-1}(t)\bar{\beta}_j(t). \eeq

As before, let us try to reduce everything to quantities which can be
numerically computed.
\begin{eqnarray*}
&& \langle H(t)\rangle+2\sum\limits_j^N \lambda_j(t) \non \\
&=& 4 \sum\limits_j^N \lambda_j(t) \bra{\Psi(t)}\bar{b}_{N+j}(t) \bar{b}_j(t)
\ket{\Psi(t)} \non \\
&=& 4 \sum\limits_{j,k,l}^N \lambda_j(t) \bra{\Psi(t)} G_{N+j,k}^{-1}(t)
\bar{\beta}_k(t) G_{j,l}^{-1}(t)\bar{\beta}_l(t) \ket{\Psi(t)} \non \\
&=& 4 \sum\limits_{j,k,l}^N \lambda_j(t) G_{N+j,k}^{-1}(t)\bra{\Psi(t)}
\bar{\beta}_k(t) \bar{\beta}_l(t) \ket{\Psi(t)} G_{j,l}^{-1}(t).
\end{eqnarray*}

Now using
\begin{eqnarray*} \bar{\beta}_j &=& \beta_j ~~{\rm for}~~ j \leq N, \non \\
\bar{\beta}_j &=& \beta^{\dagger}_j ~~{\rm for}~~ N < j \leq 2N,
\end{eqnarray*}
we have
\begin{eqnarray*}
&& \bra{\Psi(t)}\bar{\beta}_k(t)\bar{\beta}_l(t)\ket{\Psi(t)} \non \\
&=& \bra{\Psi(t)}\beta_k(t)\beta_{l}(t)\ket{\Psi}(t)=0 ~~{\rm for}~~
k\leq N ,l\leq N \non \\
&=& \bra{\Psi(t)}\beta_k(t)\beta^{\dagger}_{l-N}(t)\ket{\Psi(t)} =
\delta_{k,l-N} ~~{\rm for}~~ k\leq N, l> N, \non \\
&=& \bra{\Psi(t)}\beta^{\dagger}_{k-N}(t)\beta_l(t)\ket{\Psi(t)}=0 ~~{\rm
for}~~ k>N, l\leq N, \non \\
&=& \bra{\Psi(t)}\beta^{\dagger}_{k-N}(t)\beta^{\dagger}_{l-N}(t)\ket{\Psi(t)}
=0 ~~{\rm for}~~ k\leq N, l\leq N. \end{eqnarray*}

Hence the required expression simplifies to
\beq \langle H(t)\rangle =4 \sum\limits_{j,k}^N \lambda_j(t)
G_{N+j,k}^{-1}(t) G_{j,k+N}^{-1}(t) -2\sum\limits_j^N \lambda_j(t). \eeq
Finally the expression for the residual energy is given by
\beq E_{res}=[4 \sum\limits_{j,k}^N \lambda_j(t) G_{N+j,k}^{-1}(t)
G_{j,k+N}^{-1}(t) ]/|E_G(t)|. \eeq
$E_G$ is calculated simply by summing over all the negative eigenvalues of
the numerically diagonalized Hamiltonian, namely, $-\sum_{j=1}^N \lambda_j(t)$,
and the first term is calculated from $G$, which we already have to
calculate numerically to find the adiabatic fidelity.

\end{document}